\documentclass{article}
\usepackage[utf8]{inputenc}
\usepackage{authblk}
\usepackage{setspace,url,hyperref,graphicx}
\usepackage{amsmath,amsthm,amssymb,amsfonts,xcolor}
\usepackage{enumitem}
\usepackage[margin=0.8in]{geometry}
\usepackage[sorting = none, backend = bibtex, style=numeric-comp]{biblatex}
\usepackage{algcompatible}
\usepackage{newfloat}

\DeclareFloatingEnvironment[
    fileext=loa,
    listname=List of Algorithms,
    name=ALGORITHM,
    placement=tbhp,
]{algorithm}

\title{4-clique Network Minor Embedding for Quantum Annealers}

\author[1]{Elijah Pelofske\thanks{Email: epelofske@lanl.gov}}
\affil[1]{Los Alamos National Laboratory, CCS-3 Information Sciences}

\date{\vspace{-6ex}}

\begin{document}

\maketitle

\begin{abstract}
Quantum annealing is a quantum algorithm for computing solutions to combinatorial optimization problems. This study proposes a method for minor embedding optimization problems onto sparse quantum annealing hardware graphs called 4-clique network minor embedding. This method is in contrast to the standard minor embedding technique of using a path of linearly connected qubits in order to represent a logical variable state. The 4-clique minor embedding is possible on Pegasus graph connectivity, which is the native hardware graph for some of the current D-Wave quantum annealers. The Pegasus hardware graph contains many cliques of size 4, making it possible to form a graph composed entirely of paths of connected 4-cliques on which a problem can be minor embedded. The 4-clique chains come at the cost of additional qubit usage on the hardware graph, but they allow for stronger coupling within each chain thereby increasing chain integrity, reducing chain breaks, and allow for greater usage of the available energy scale for programming logical problem coefficients on current quantum annealers. The 4-clique minor embedding technique is compared against the standard linear path minor embedding with experiments on two D-Wave quantum annealing processors with Pegasus hardware graphs. We show proof of concept experiments where the 4-clique minor embeddings can use weak chain strengths while successfully carrying out the computation of minimizing random all-to-all spin glass problem instances. 

\end{abstract}

\section{Introduction}
\label{section:introduction}

Quantum annealing is a type of analog quantum computation which is effectively a relaxed version of Adiabatic Quantum Computing (AQC). Quantum annealing is designed to solve combinatorial optimization problems by using quantum fluctuations in order to minimize an encoded problem Hamiltonian~\cite{https://doi.org/10.48550/arxiv.quant-ph/0001106, morita2008mathematical, das2008colloquium, Hauke_2020, PhysRevX.4.021041, santoro2006optimization, johnson2011quantum, boixo2013experimental}. This process of computation works by starting the system in the easy-to-prepare ground state of a Hamiltonian, and then slowly transitioning the system into a second Hamiltonian which we wish to find the ground state of (generally this second Hamiltonian corresponds to a problem which we do not know the ground state of because it is difficult to compute). In the transverse field Ising model version of quantum annealing, the system is put into an initial uniform superposition that is the ground state of the Hamiltonian:

\begin{equation}
    H_{initial} = \sum_{i}^n \sigma_i^x
    \label{equation:transverse_ising}
\end{equation}

Where $\sigma_i^x$ is the Pauli matrix for each qubit at index $i$. The user programmed problem Hamiltonian is then turned on over the course of the anneal:

\begin{equation}
    H(t) = A(t)H_{initial} + B(t)H_{ising}
    \label{equation:QA_eq}
\end{equation}

Combined, $A(t)$ and $B(t)$ define the \emph{anneal schedules}. Typically, at $t=0$ the $A(t)$ term is dominating as the system is prepared in the ground state of $H_{initial}$, and therefore the qubits are put into an initial superposition, and at the end of the anneal $B(t)$ (the problem Hamiltonian) is dominating. The \emph{annealing time} over which these schedules are applied can be varied, and in the case of physically implemented quantum annealers the possible annealing time is constrained by the hardware. At the end of the anneal, the qubit states are read out as classical bits, which correspond to the variable states. Those samples are intended to be low energy solutions to the problem Ising $H_{ising}$ that the user has specified. For D-Wave quantum annealers, the user can program the anneal schedule - specifically the ratio between $A(s)$ and $B(s)$ (known as the anneal fraction) for each point in time during the anneal. The classical problem Hamiltonian, composed of single and two body terms, is defined as:

\begin{equation}
    H_{ising} = \sum_{i}^n h_i \sigma_i^z + \sum_{i < j}^n J_{ij} \sigma_i^z \sigma_j^z
    \label{equation:problem_ising}
\end{equation}

The classical problem Hamiltonian is equivalent to large class of combinatorial optimization problems. Specifically for optimization problems where the decision variables are discrete; for Ising models the variable states can be either $+1$ or $-1$. There are formulations known for converting many NP-Hard problems into discrete combinatorial optimization problems of this form, thereby allowing quantum annealers to sample low energy classical solutions of the programmed optimization problems. The company D-Wave has created a number of quantum annealing processors using superconducting flux qubits; these devices have been applied to a wide range of problems~\cite{PhysRevX.5.031040, 9259934, Hauke_2020}. Quantum annealing has experimentally been shown to be competitive for good heuristic sampling of combinatorial optimization problems~\cite{https://doi.org/10.48550/arxiv.2210.04291, bauza2024scaling, Albash_2018, Pelofske_2022} and simulation of frustrated magnetic systems~\cite{King_2021, King_2021_spin_ice, Lopez_Bezanilla_2023}. Although the size (e.g. the number of qubits) has been increasing as hardware development improves, the hardware of these quantum annealers is relatively sparsely connected, this limiting what problem connectivity graphs can be sampled on these devices.

Minor embedding is the only mechanism that allows logical problems with structure that does not directly match the underlying hardware graph to be programmed onto the hardware~\cite{klymko2014adiabatic, bernal2020integer, choi2011minor, choi2008minor, boothby2016fast, PRXQuantum.2.040322, zbinden2020embedding, lucas2019hard}. In minor embedding for quantum annealing, each variable on the logical problem graph can be represented by a collection of physical qubits which are linked together ferromagnetically. The ferromagnetic coupling attempts to ensure that the variables representing each logical qubit are in agreement as to what the logical variable state is; there is an energy penalty for a qubit not having the same spin state as its neighbors. The standard minor embedding that is used creates a linear path of physically linked qubits, typically a linear nearest neighbors (LNN) graph, which form this ferromagnetically-bound logical variable. These linear path groups of qubits are typically referred to as \emph{chains}. 

The essential idea of computing graph minors, in particular for embedding a problem onto a fixed hardware graph, is that the minor does not need to be a linear path; it can be effectively any graph structure which is embeddable onto the target graph. The constraint is that the minor embeddings ideally should require as little additional hardware qubits (and couplers) as possible to fit more problems, or larger problems, onto the hardware chip. Therefore, the linear paths are used so as to reduce the overhead of using additional physical qubits, so that each logical variable can be routed so that the required logical quadratic variable interactions can occur. 

\begin{figure}[h!]
    \centering
    \includegraphics[width=0.3\textwidth]{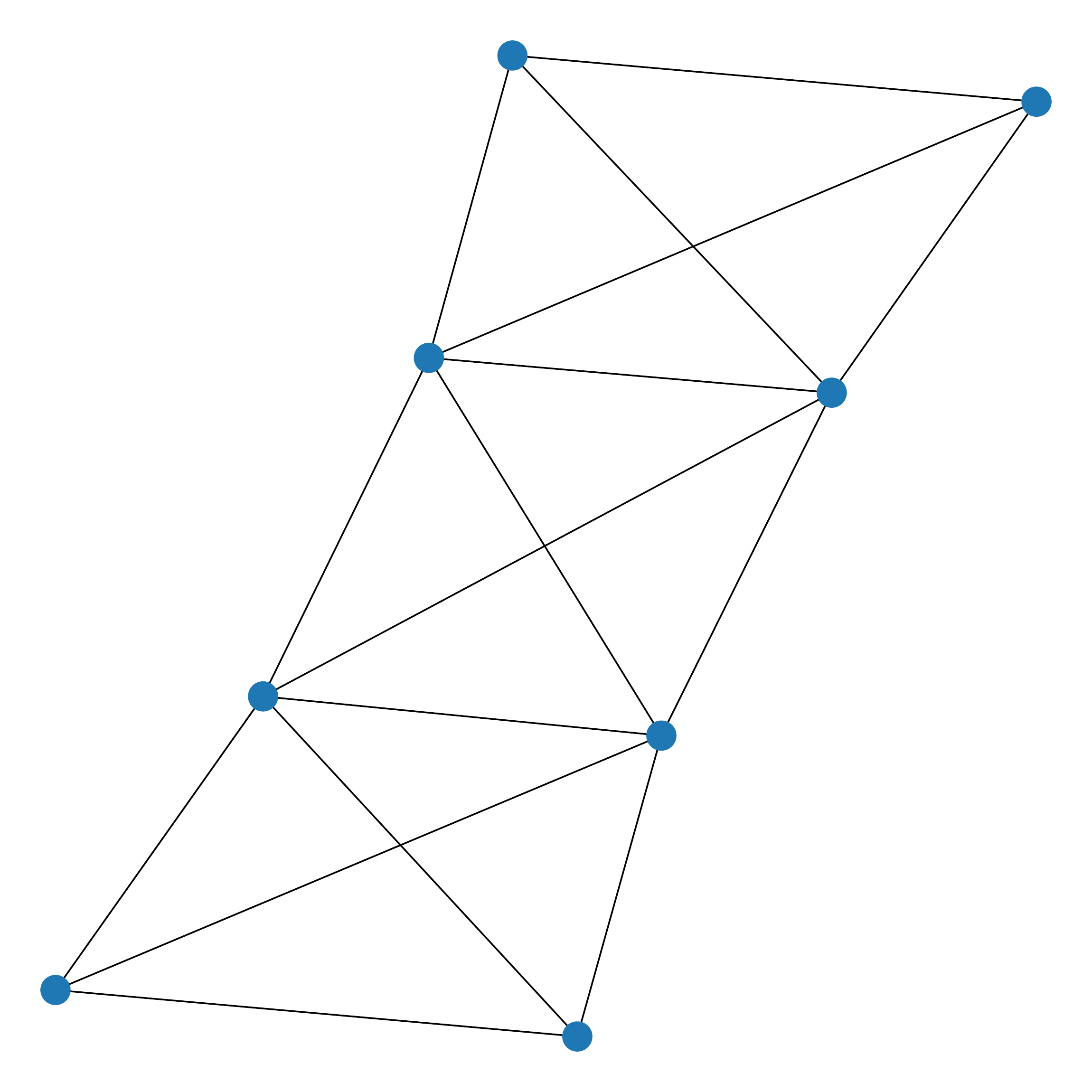}
    \caption{A path of 4-cliques. }
    \label{fig:3_4_cliques}
\end{figure}

In this article we propose a new method of minor embedding which we will refer to as \emph{4-clique network minor embedding}. This method is based on a property of one of the D-Wave quantum annealing device connectivities, called Pegasus~\cite{https://doi.org/10.48550/arxiv.2003.00133, https://doi.org/10.48550/arxiv.1901.07636, boothby2021architectural, zbinden2020embedding}, which (while still quite sparse) contains a large number of 4-cliques throughout its hardware lattice. The first D-Wave devices that were manufactured had a connectivity graph called Chimera~\cite{zbinden2020embedding, Lobe_2021}, which is sparser than Pegasus. The newest generation of D-Wave quantum annealing hardware has a graph connectivity called Zephyr~\cite{boothby2021architectural, zephyr}. With the 4-cliques in the Pegasus hardware graph, it is possible to form connected paths of 4-cliques (see Figure~\ref{fig:3_4_cliques}) from \emph{4-clique chains} in order to create minor embeddings onto Pegasus chip hardware. While the 4-clique minor embedding uses more qubits, it allows significantly more ferromagnetic couplers to be programmed in each chain, which reinforces the integrity of each chain more than a linear path connectivity. With linear path embeddings the measured qubits in a chain often disagree on their logical spin state, especially for very long chains. Chains that have measured spins which are not the same are referred to as a \emph{broken chains}. Therefore, 4-clique embeddings allow more ferromagnetic chain break penalty weights to be used for each logical variable, thereby ideally reducing the number of chain breaks. The intuitive reasoning is that with a greater number of ferromagnetic couplers enforcing the state of each logical variable, there will be fewer chain breaks and therefore smaller chain strengths can be used. 

One of the problems in general with minor embedding is that the strong ferromagnetic couplers can dominate the programmable energy range on the chip; current D-Wave devices have a set range of physical weights that the user can program to specify the problem Hamiltonian. There is limited precision when encoding these weights (approximately two decimal places), and therefore it is important to use as much of the physical programmable weight range as possible for the problem coefficients (instead of using those weights for encoding a minor embedding). Adding in the strong ferromagnetic chain couplers reduces the effective range that can be used to program the logical problem weights. The 4-clique network minor embedding is therefore useful for a second reason; that is by increasing the number of ferromagnetic couplers per chain, the (relative) magnitude of the ferromagnetic couplers can be reduced compared to linear path embeddings which could allow for a larger effective programming weight range to be used for programming the weights of the logical problem. 

\begin{table*}[t!]
\begin{center}
\begin{tabular}{ |p{4cm}||p{2cm}|p{1.4cm}|p{1.4cm}|p{2.4cm}| } 
 \hline
 D-Wave QPU chip id & Topology name & Available qubits & Available couplers & Annealing time (min, max) \newline [microseconds] \\ 
 \hline
 \hline
 \texttt{Advantage\_system4.1} & Pegasus $P_{16}$ & 5627 & 40279 & (0.5, 2000) \\ 
 \hline
 \texttt{Advantage\_system6.1} & Pegasus $P_{16}$ & 5616 & 40135 & (0.5, 2000) \\ 
 \hline
\end{tabular}
\end{center}
    \caption{D-Wave quantum annealing processor hardware summary. Note that each of these devices have some hardware defects which cause the available hardware (qubits and couplers) to be smaller than the ideal Pegasus $P_{16}$ graph lattice structure. }
    \label{table:hardware_summary}
\end{table*}

\begin{algorithm*}[t!]
\begin{algorithmic}[1]
    \STATE{\textbf{Input:} Hardware graph $G$}
    \STATE{\texttt{4-cliques} $\leftarrow$ Compute all cliques of size $4$ in $G$}
    \FOR{$K \in \texttt{4-cliques}$}
        \STATE{If any of the nodes in $K$ have already been contracted in a previous iteration, skip this iteration}
        \STATE{Randomly choose two of the nodes $n_1, n_2$ from $K$}
        \STATE{Contract edge ($n_1$, $n_2$) to form a node called $n_1 n_2$}
        \STATE{Choose the remaining two nodes $n_3, n_4$ from $K$}
        \STATE{Contract edge ($n_3$, $n_4$) to form a node called $n_3 n_4$}
        \STATE{Remove any self edges that may have been generated from these two edge contractions}
    \ENDFOR
    \STATE{Remove all nodes and edges in $G$ which were not formed by edge contraction. Specifically, remove all nodes that are not named with the form $n_x n_y$ and remove all edges that were not formed by edge contractions (e.g. remove all edges that do not have the form of $(n_{x_1} n_{y_1}, n_{x_2} n_{y_2})$ )}
    \STATE{\textbf{Return:} $G$}
\end{algorithmic}
\caption{Contract hardware graph to a 4-clique network}
\label{algorithm:contract_hardware_graph}
\end{algorithm*}

We note that a related idea is Quantum Annealing Correction (QAC)~\cite{pudenz2015quantum, vinci2015quantum, vinci2016nested, mishra2016performance, pudenz2014error, bauza2024scaling, Matsuura_2019, Matsuura_2017}, where the states of problem variables are reinforced using ferromagnetic couplings to a common penalty qubit.

Section~\ref{section:methods} describes the 4-clique graph construction in detail, along with creating some example minor embeddings. Section~\ref{section:results} describes the quantum annealing experimental results on the 4-clique minor embeddings compared to linear path minor embeddings when executed on D-Wave quantum annealers. Section~\ref{section:conclusion} concludes with what the results show in regards the effectiveness of the 4-clique minor embedding and future research questions. The figures in this article were generated using matplotlib~\cite{thomas_a_caswell_2021_5194481, Hunter:2007}, networkx~\cite{hagberg2008exploring}, and dwave-networkx~\cite{DWave_networkx} in Python 3. Data associated with this paper, including raw D-Wave measurements and minor embeddings, is available as a public dataset~\cite{elijah_pelofske_2023_7552776}.

\section{Methods}
\label{section:methods}

Section~\ref{section:methods_4_clique_minor_embedding} describes the 4-clique graph construction on a Pegasus graph, and how the minor embedding process works using this 4-clique network. Section~\ref{section:methods_implementation} describes the implementation of the 4-clique minor embeddings on quantum annealing hardware; specifically the problem instances which are used to compare the 4-clique and equivalent linear path minor embeddings are described, along with the D-Wave parameter settings used for the experiments.

\subsection{4-clique network minor embedding}
\label{section:methods_4_clique_minor_embedding}

\begin{figure*}[t!]
    \centering
    \includegraphics[width=0.32\textwidth]{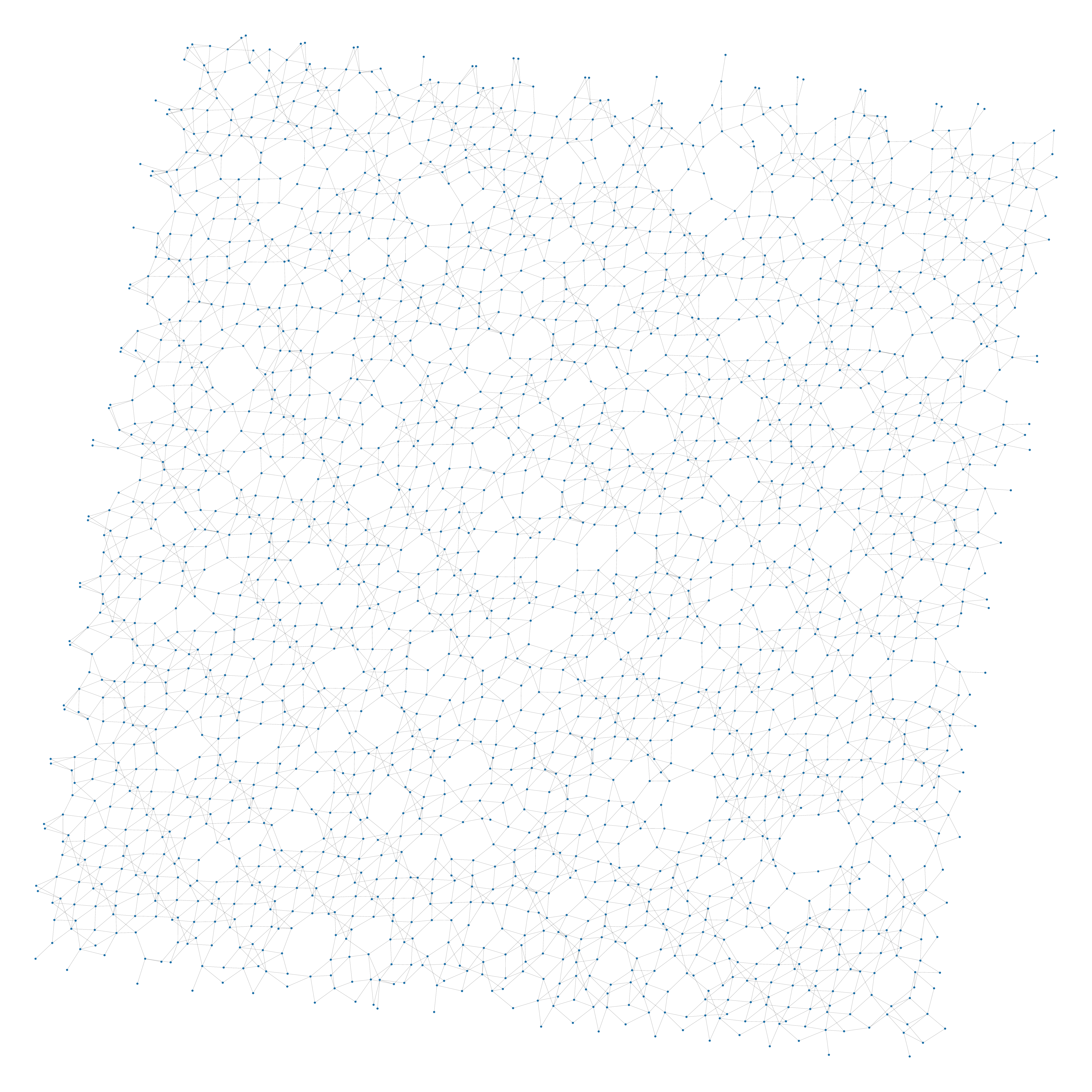}
    \includegraphics[width=0.32\textwidth]{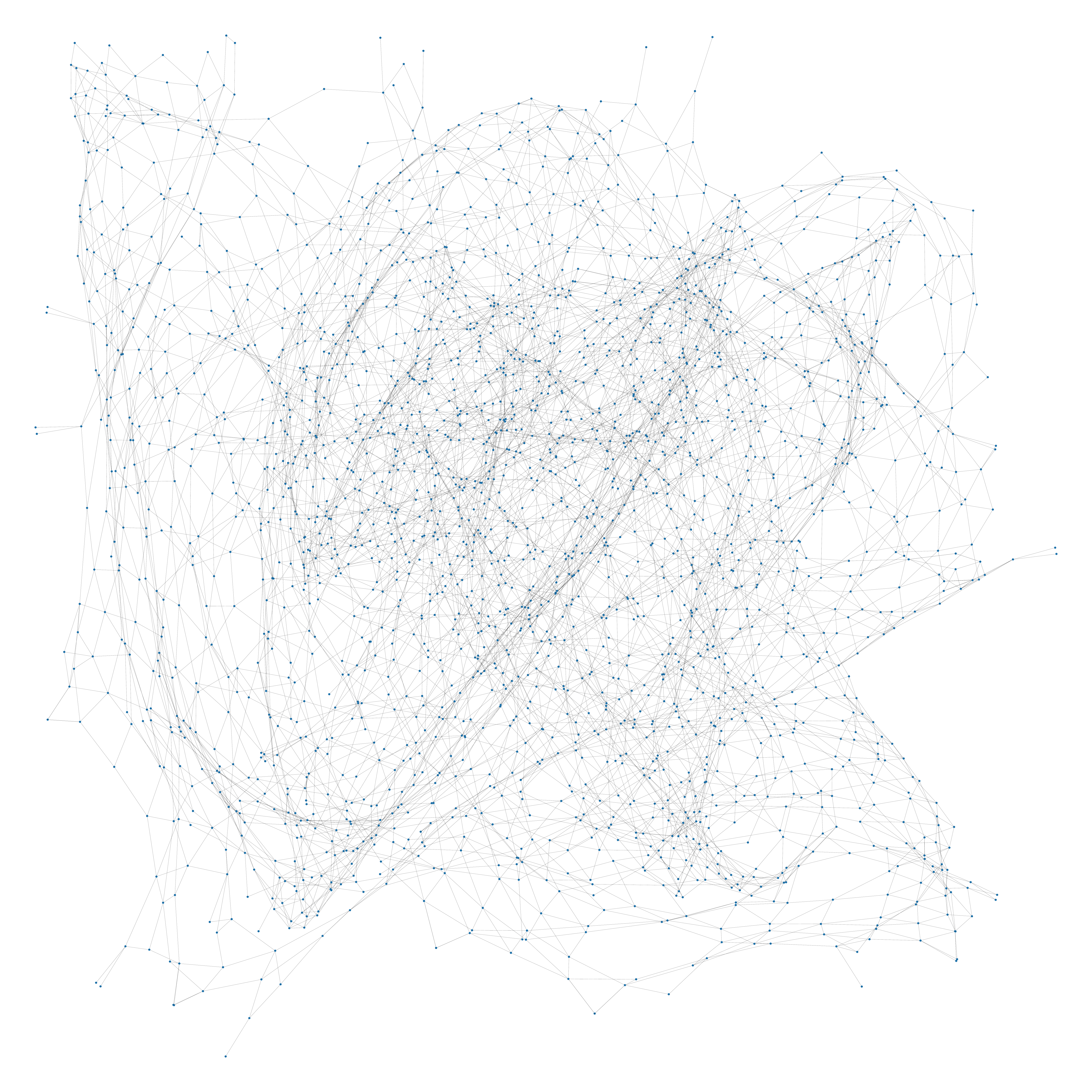}
    \includegraphics[width=0.32\textwidth]{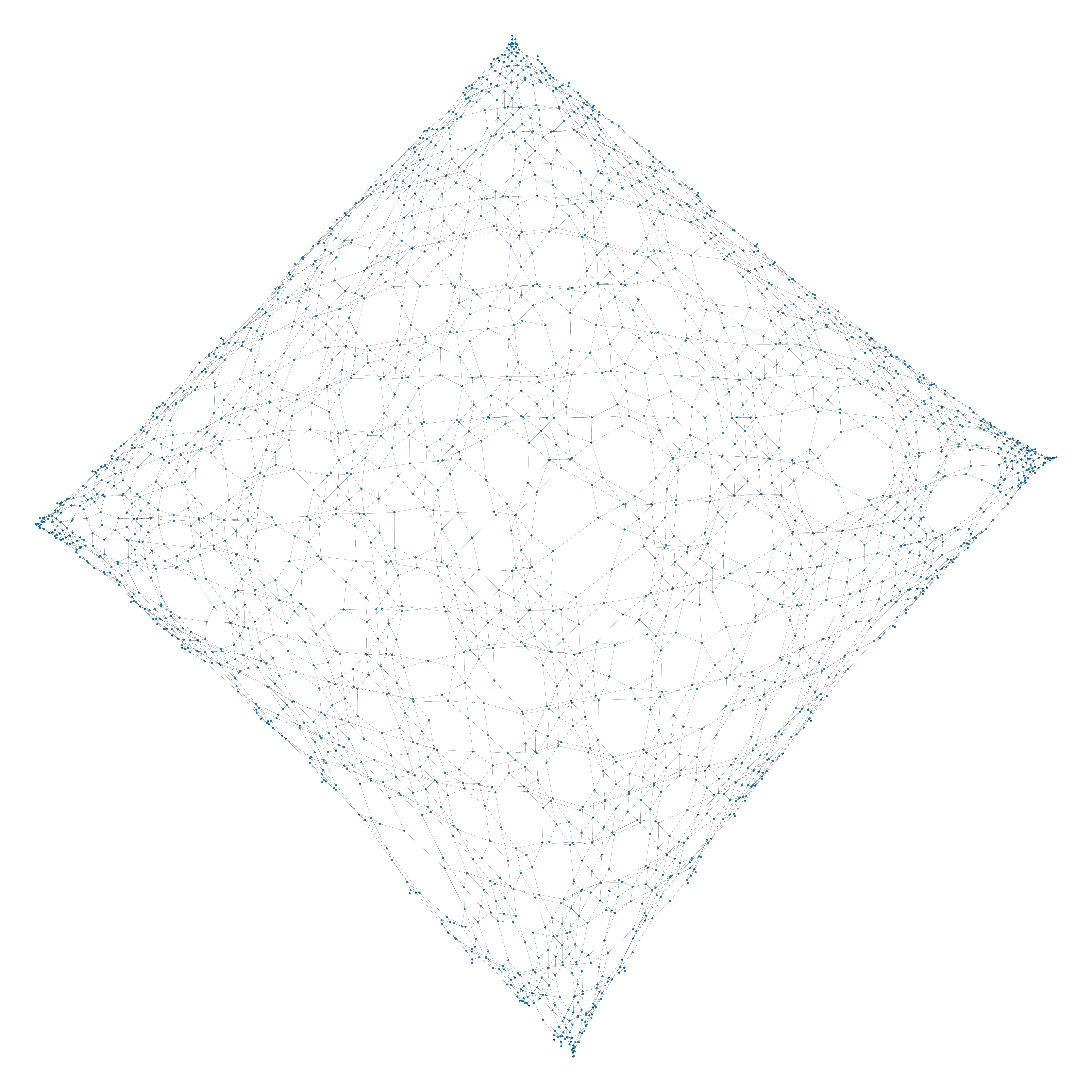}\\
    \includegraphics[width=0.32\textwidth]{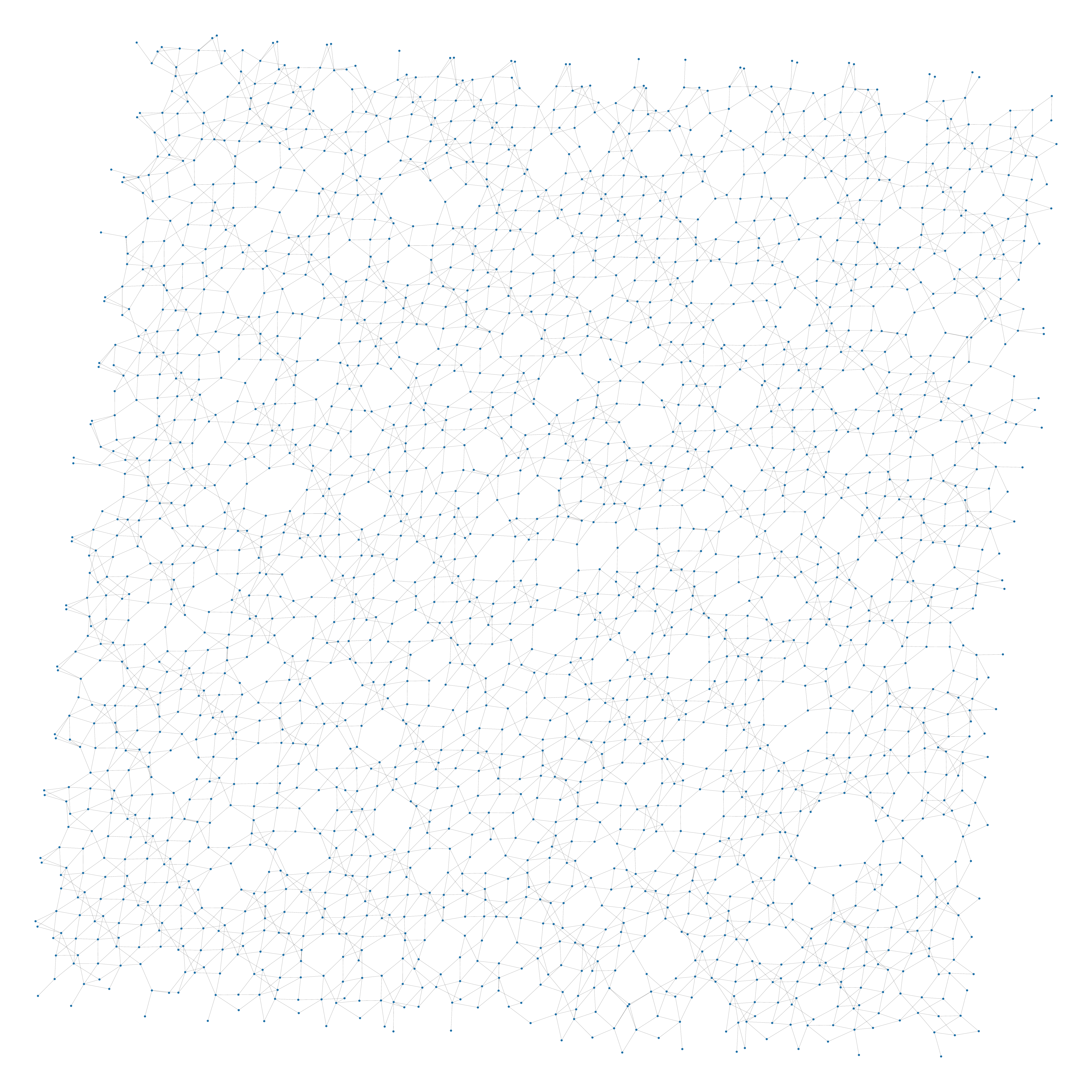}
    \includegraphics[width=0.32\textwidth]{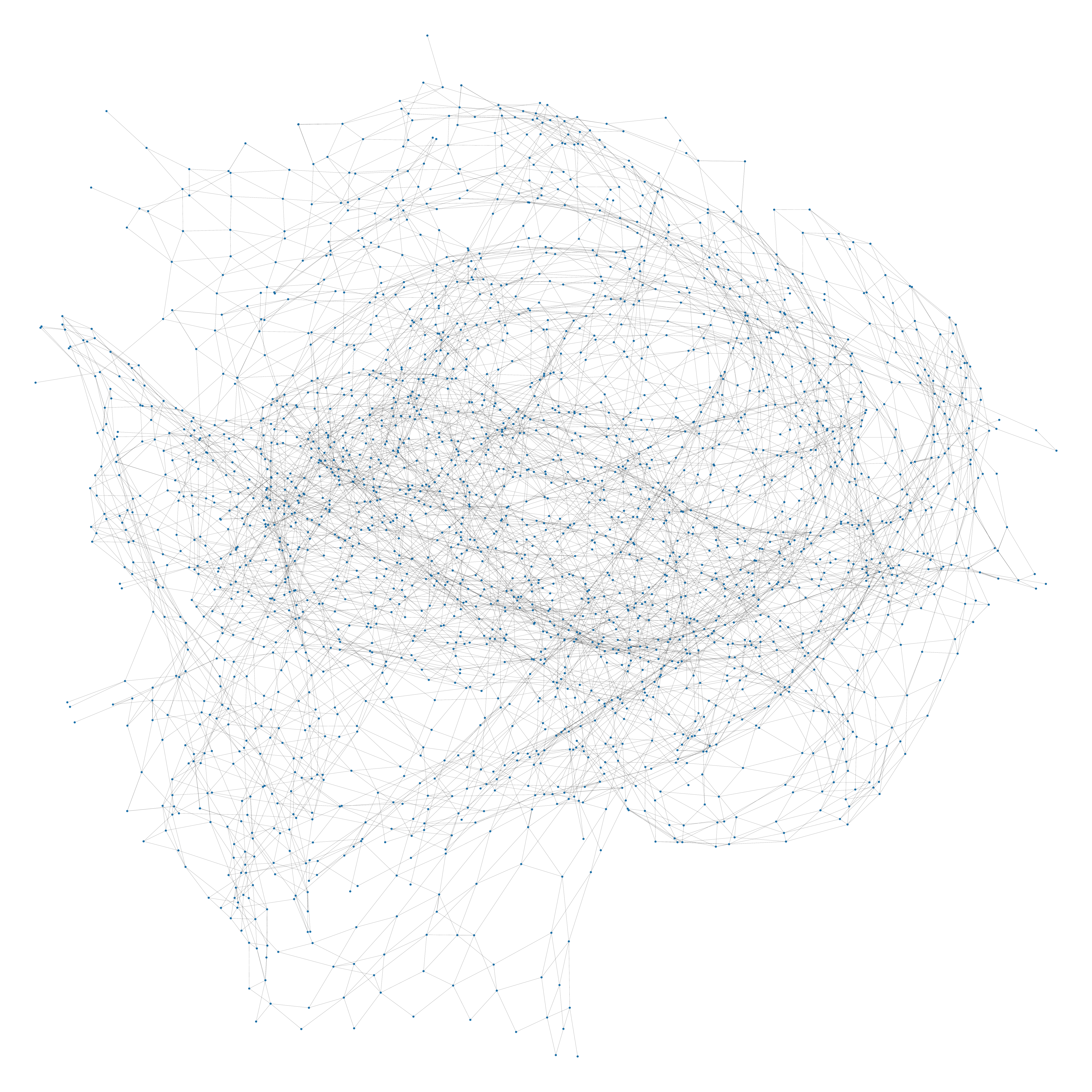}
    \includegraphics[width=0.32\textwidth]{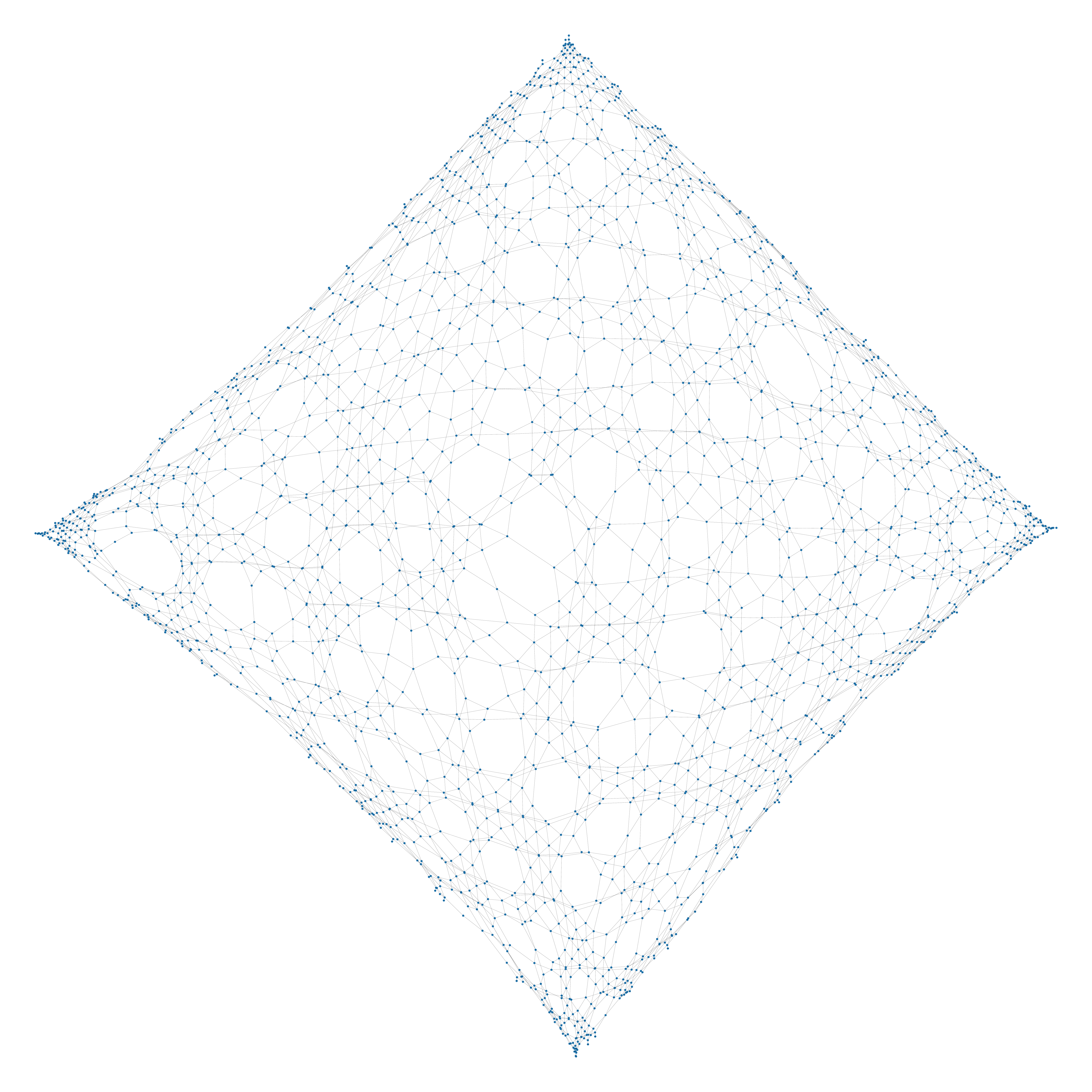}
    \caption{Contracted 4-clique graph of \texttt{Advantage\_system4.1} with kamada kawai layout (top left), spring layout (top middle), and spectral layout (top right). Contracted 4-clique graph of \texttt{Advantage\_system6.1} with kamada kawai layout (bottom left), spring layout (bottom middle), and spectral layout (bottom right). The contracted 4-clique graph of \texttt{Advantage\_system4.1} has $2471$ nodes and $6270$ edges. The contracted 4-clique graph of \texttt{Advantage\_system6.1} has $2463$ nodes and $6245$ edges. As defined by Algorithm~\ref{algorithm:contract_hardware_graph}, an edge in a contracted 4-clique graph represents $4$ edges in the underlying hardware graph, and a node represents $2$ physical qubits in the hardware graph. These contracted clique graphs are quite sparse; the maximum clique of both of these graphs are $2$. When the clique contraction is performed on the hardware graphs, there are many small unconnected components that are generated. These figures are showing only the largest connected component since it is the one which can be used for large minor embeddings.  }
    \label{fig:contracted_cliques}
\end{figure*}

Algorithm~\ref{algorithm:contract_hardware_graph} constructs a network of 4-clique paths from a hardware graph connectivity. Algorithm~\ref{algorithm:contract_hardware_graph} assumes that the hardware graph contains at least one clique of size $4$, otherwise it does not contract any edges in the graph all. For example, Algorithm~\ref{algorithm:contract_hardware_graph} performs no contractions to a Chimera graph since the Maximum Clique of a Chimera lattice is $2$. The largest connected component (there are other smaller components which are not connected to the main graph) of the contracted 4-clique graphs of \texttt{Advantage\_system6.1} and \texttt{Advantage\_system4.1} are shown using different layout algorithms in Figure~\ref{fig:contracted_cliques}. Note that the resulting contracted 4-clique graph from Algorithm~\ref{algorithm:contract_hardware_graph} may not be connected, and may not have large connected components. The set of connected components, and thus the largest connected component, was computed using networkx~\cite{Networkx_connected_components, hagberg2008exploring}. Once this contracted 4-clique graph of a target hardware graph has been created, standard minor embedding algorithms such as minorminer~\cite{https://doi.org/10.48550/arxiv.1406.2741, DWave_github_minorminer} can be applied to create a minor embedding composed of chains of 4-cliques, such as in Figure~\ref{fig:3_4_cliques}. For the purpose of constructing the 4-clique graph from the contracted clique graph provided by Algorithm~\ref{algorithm:contract_hardware_graph}, one can take the subgraph of the device hardware graph induced by the separated nodes $n_x$, $n_y$, given by the name of the contracted nodes for all of the contracted nodes in the 4-clique chain. Table~\ref{table:hardware_summary} shows a hardware summary in terms of qubits and couplers for the two Pegasus hardware graph D-Wave devices that are used for analyzing and implementing example spin glass problems, embedded using the 4-clique network minor embeddings (and compared against the linear path minor embeddings). The largest connected component of the contracted 4-clique network for \texttt{Advantage\_system4.1} uses a proportion of $0.878$ of the available hardware qubits, and a proportion of $0.623$ of the available hardware couplers. The largest connected component of the contracted 4-clique network for \texttt{Advantage\_system6.1} uses a proportion of $0.877$ of the available hardware qubits, and a proportion of $0.622$ of the available hardware couplers. 

For making two variable interactions possible in the 4-clique minor embedding, since all chains are part of a 4-clique network already, every 4-clique chain can be connected to another 4-clique chain by 4 physical couplers for each single coupler that was computed in the minor embedding for the contracted clique graph. Note that although a linear path minor embedding is usually computed by forming a single path of connected qubits, it is possible for the minor embedding algorithm to use some branching if required; and therefore sometimes the standard minor embeddings are not strictly linear path (although almost always they do create linear path chains). 

\begin{figure*}[t!]
    \centering
    \includegraphics[width=0.49\textwidth]{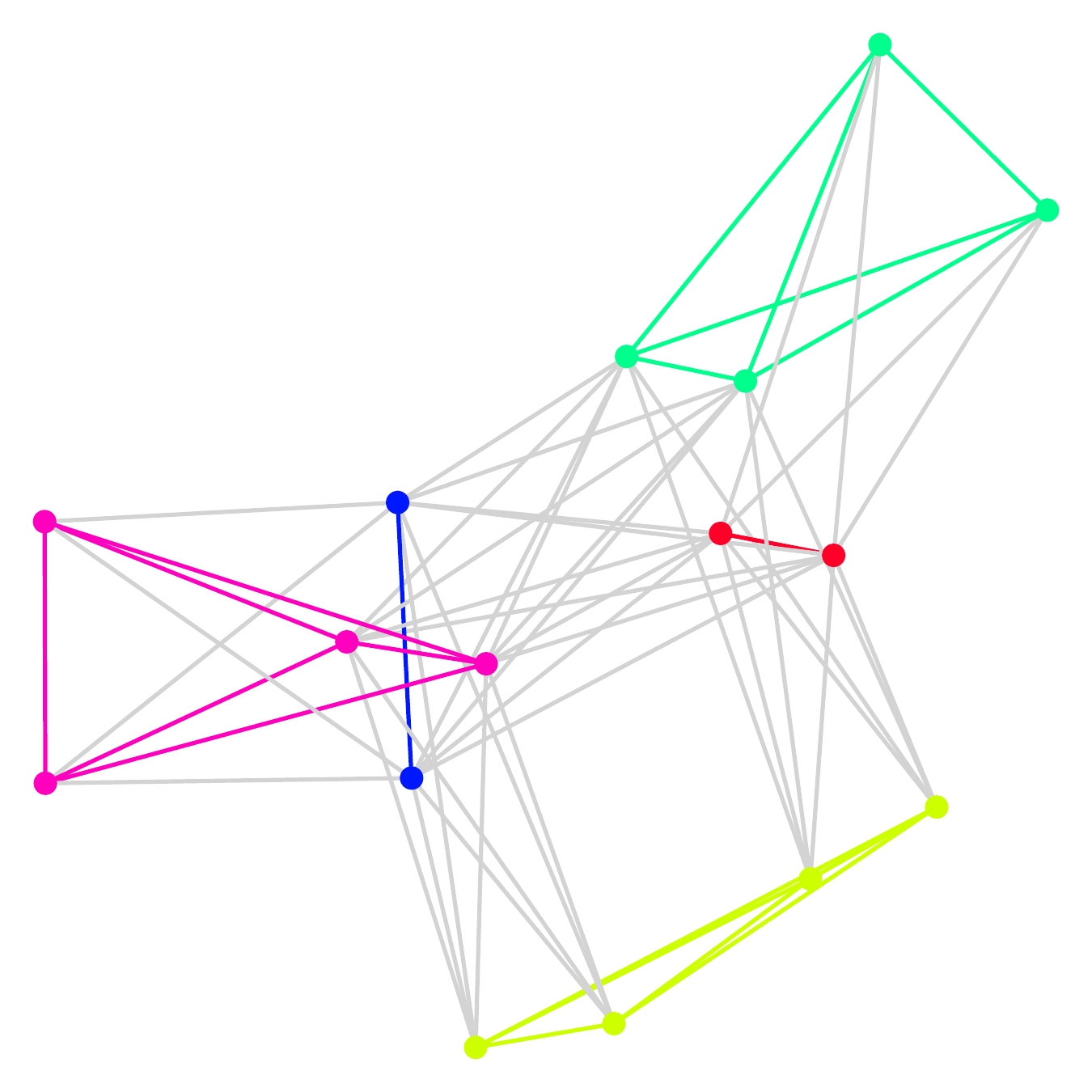}
    \includegraphics[width=0.49\textwidth]{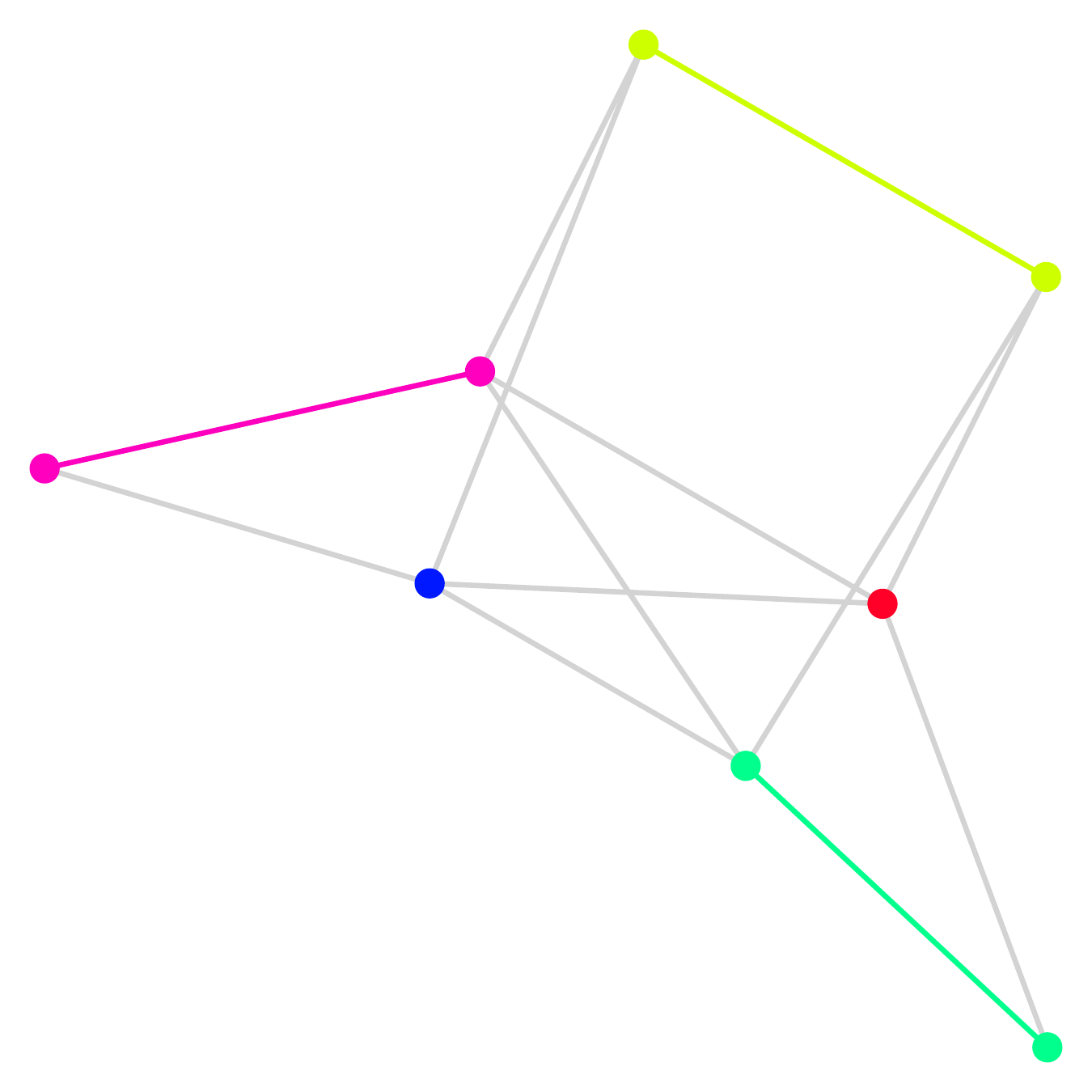}
    \caption{Minor-embedding of a $K_5$ clique onto the target graph connectivity of Pegasus, using a 4-clique minor embedding (left) and an equivalent linear path minor embedding (right). Nodes in the graph are physical qubits, and edges are physical couplers. Grey edges represent the physical couplers onto which the problem specific coefficients would be encoded. Colored edges and nodes denote the minor embeddings; each chain (comprised of qubits and couplers) is uniquely colored. Notice that the 4-clique minor embedding by default uses chains with length 2 as the smallest possible chains to encode a logical variable; these together form a single node pair $n_x$, $n_y$ that are two of the variables in a 4-clique in the hardware. The linear path embedding by contrast does use chains of length $1$ (meaning there is no chain). Qubits and couplers not used by the minor embedding are not shown. Because Pegasus natively has cliques of size 4, it makes no sense to actually use a minor-embedding of size 4. Therefore, this specific minor embedding diagram is purely for the purposes of describing the 4-clique minor embedding algorithm. }
    \label{fig:minor_embedding_N4}
\end{figure*}

\begin{figure*}[t!]
    \centering
    \includegraphics[width=0.49\textwidth]{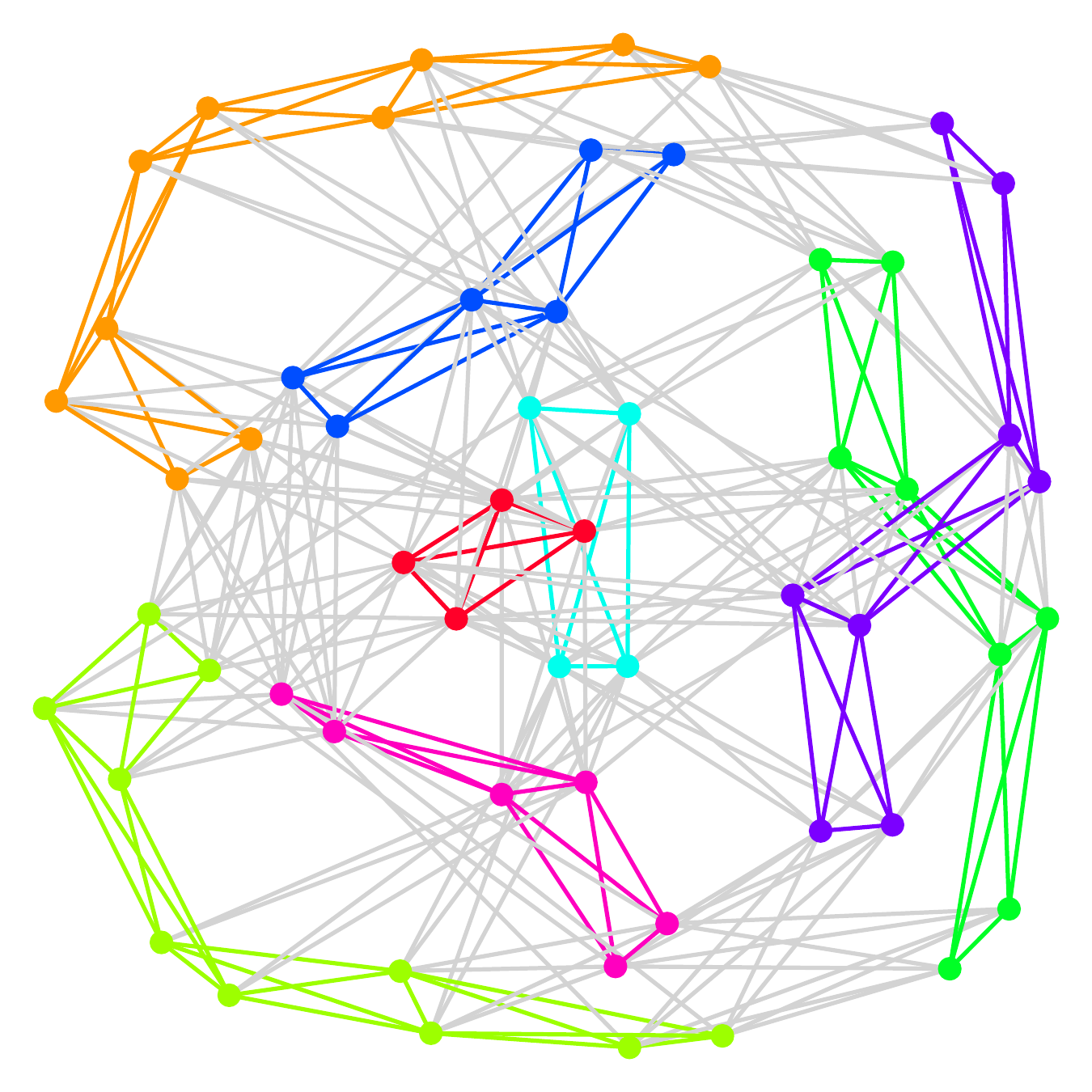}
    \includegraphics[width=0.49\textwidth]{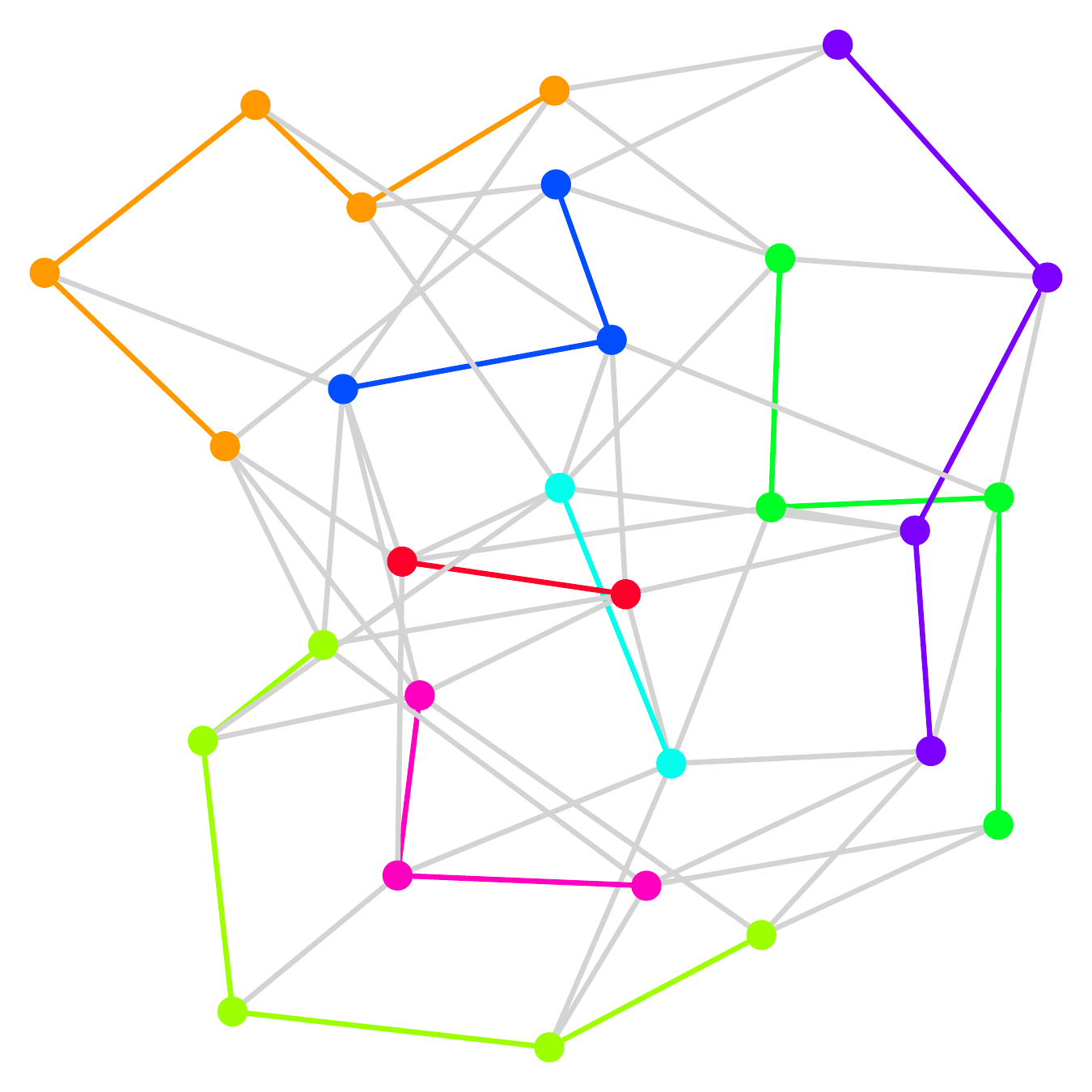}
    \caption{Minor-embedding of a $K_8$ clique using a 4-clique minor embedding (left) and an equivalent linear path minor embedding (right) on a Pegasus graph. Grey edges represent the physical couplers onto which the problem specific coefficients would be encoded. Colored edges and nodes denote the minor embeddings; each chain (comprised of qubits and couplers) is uniquely colored. As in Figure~\ref{fig:minor_embedding_N4}, the parts of the Pegasus graph which are not used by the minor embeddings are not shown. }
    \label{fig:minor_embedding_N8}
\end{figure*}

To provide a direct comparison between a 4-clique embedding and a linear path embedding, we can take any minor-embedding of a problem connectivity with the target of the contracted clique graphs (for example in Figure~\ref{fig:contracted_cliques}), and separate out a linear path path by taking one of the two variables in each node pair $n_x$, $n_y$. This means that we can compute a minor embedding of a problem on a contracted 4-clique graph, and then compute a linear path embedding of the same path length (but half the number of qubits in each chain) for the purpose of directly comparing the two embeddings since they have high hardware overlap. 

In linear path minor embedding chains, the node degrees are usually either $2$, for nodes within the chain, or $1$ for nodes at the end of the chain assuming the chain is strictly linear nearest neighbors (LNN). The minor embedding may utilize branching of a linear path, resulting in node degrees of $3$, for example. As shown by Figure~\ref{fig:3_4_cliques} in 4-clique chains the node degrees are $5$ for nodes within the chain and degree $3$ for nodes at the end of the chains (again, assuming the 4-clique path is linear, and not branching). 

\begin{figure*}[t!]
    \centering
    \includegraphics[width=0.49\textwidth]{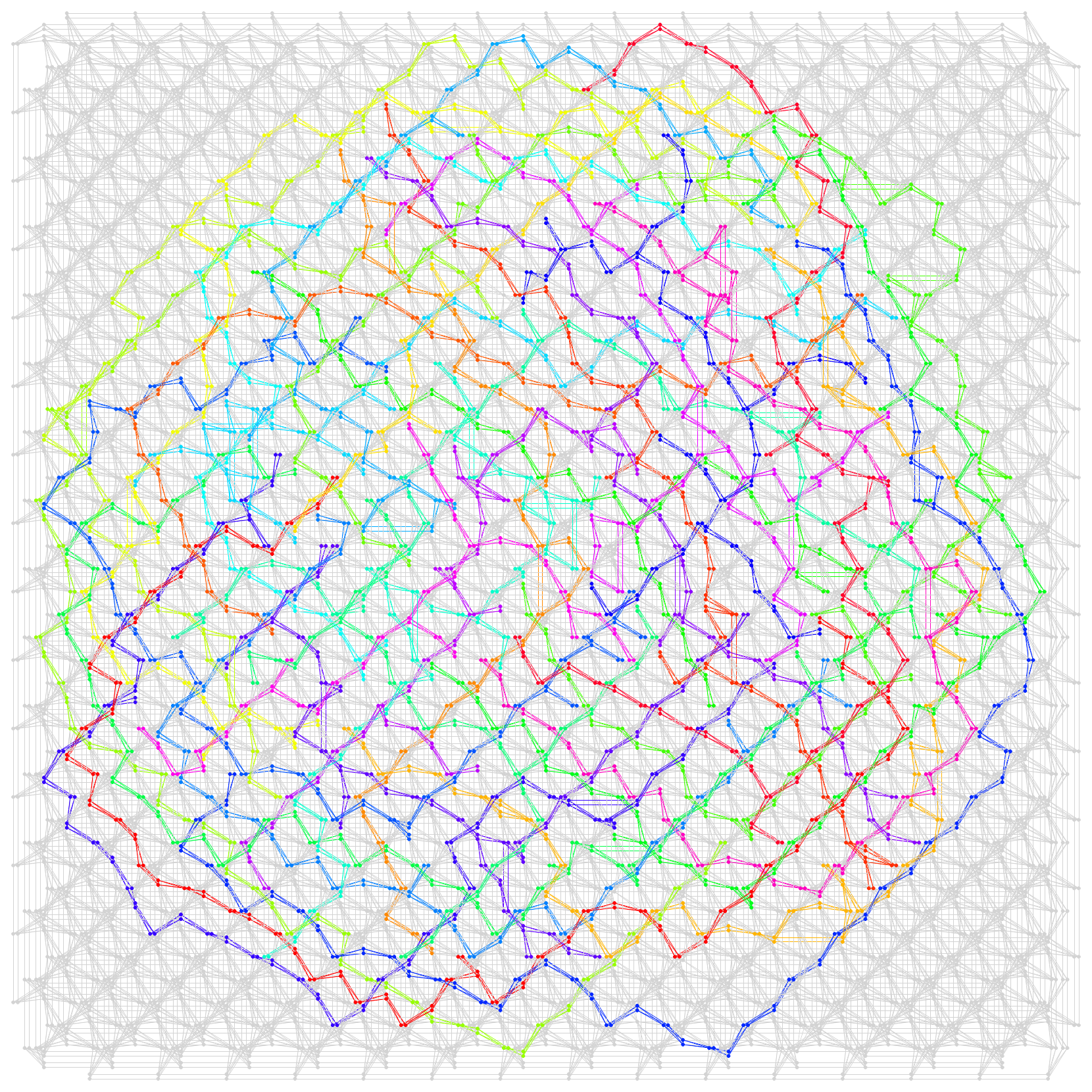}
    \includegraphics[width=0.49\textwidth]{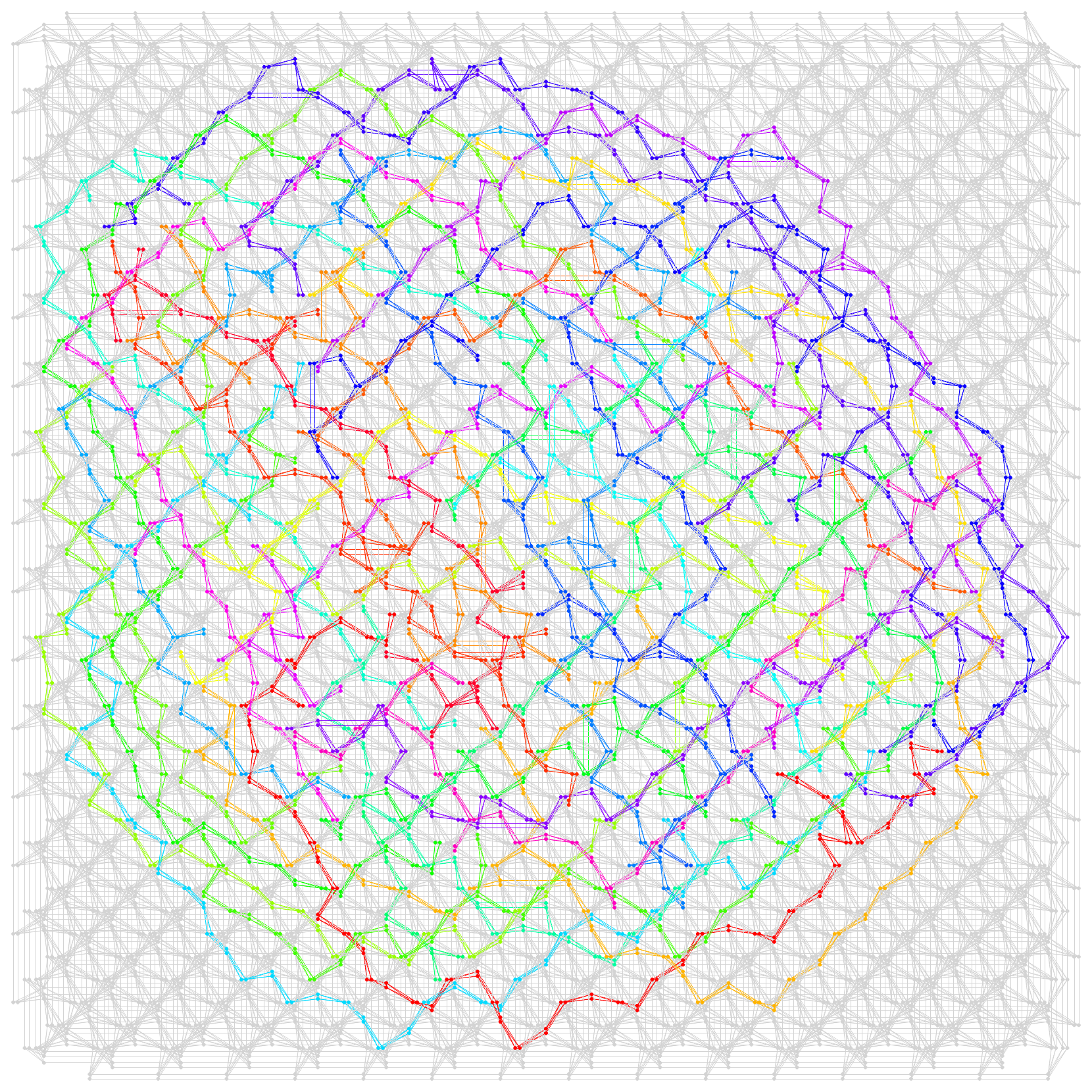}
    \caption{$K_{32}$ 4-clique minor embedding on the $P_{16}$ Pegasus hardware graphs of \texttt{Advantage\_system4.1} (left) and \texttt{Advantage\_system6.1} (right). Each of the $32$ chains are uniquely colored. These are the largest all-to-all 4-clique minor embeddings that could be computed using minorminer in a reasonable amount of time. }
    \label{fig:K32_clique_4_minor_embedding_Pegasus}
\end{figure*}

The largest all-to-all minor embeddings which could be constructed, using heuristic minor embedding algorithms, on the contracted 4-clique networks for both \texttt{Advantage\_system6.1} and \texttt{Advantage\_system4.1} are $32$ node cliques. Example 4-clique random minor embeddings for $K_{32}$ graphs are shown in Figure~\ref{fig:K32_clique_4_minor_embedding_Pegasus}, overlayed onto the Pegasus hardware graph. Table~\ref{table:minor_embedding_chain_lengths} in Appendix~\ref{section:table_4_clique_chain_lengths} details all of the computed minor embedding chain length statistics for the 4-clique minor embeddings from $N=3, \dots, 32$. 

Currently, D-Wave quantum annealers have three distinct hardware connectivities, with the names of Chimera, Pegasus, and Zephyr. Chimera graphs are too sparse for the 4-clique minor embedding scheme (the maximum clique of a Chimera graph is $2$). Pegasus works perfectly for this idea; in fact its highly connected 4-cliques motivated this idea. Zephyr also has cliques of size $4$, however Zephyr can not form a large fully connected 4-clique network, instead the 4-clique contractions result in disconnected subgraphs. The contracted clique graph of a Zephyr $Z_{16}$ graph is shown in Appendix~\ref{section:4_clique_zephyr}. This means that large 4-clique minor embeddings could not be created using Zephyr hardware, whereas for Pegasus larger 4-clique minor embeddings can be created.

\begin{figure*}[t!]
    \centering
    \includegraphics[width=0.49\textwidth]{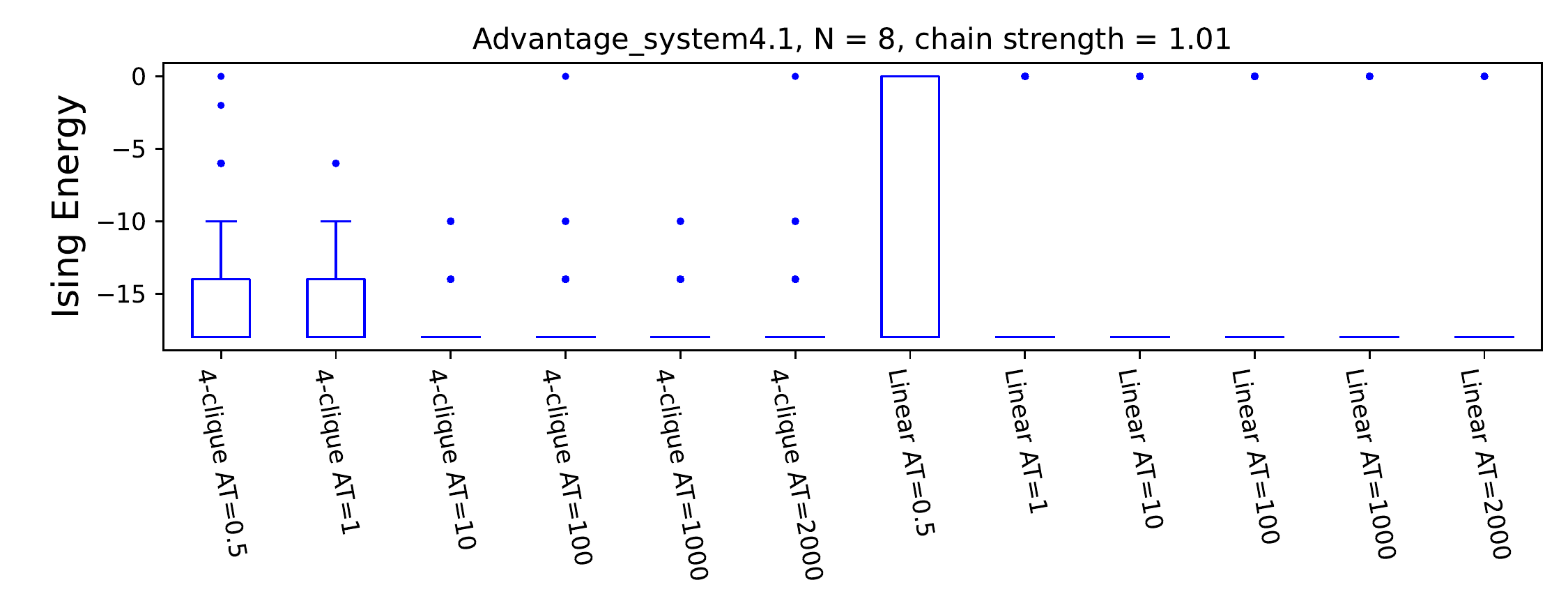}
    \includegraphics[width=0.49\textwidth]{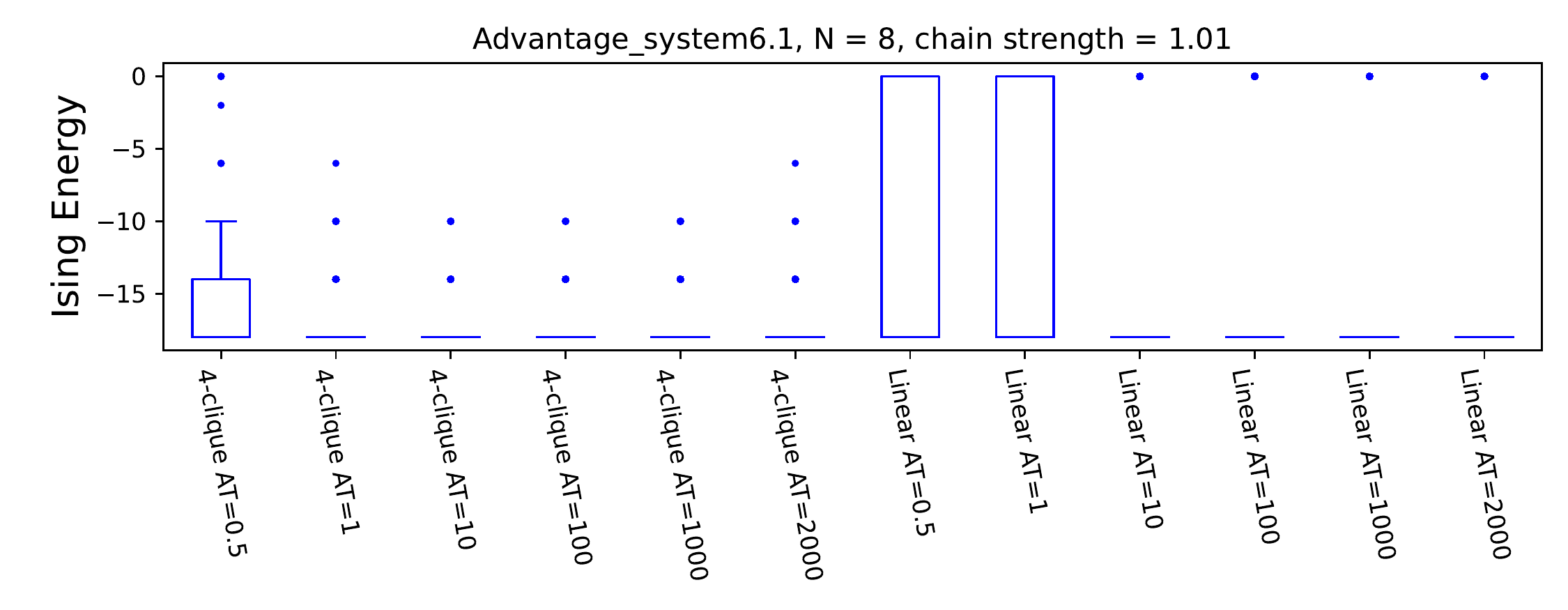}
    \includegraphics[width=0.49\textwidth]{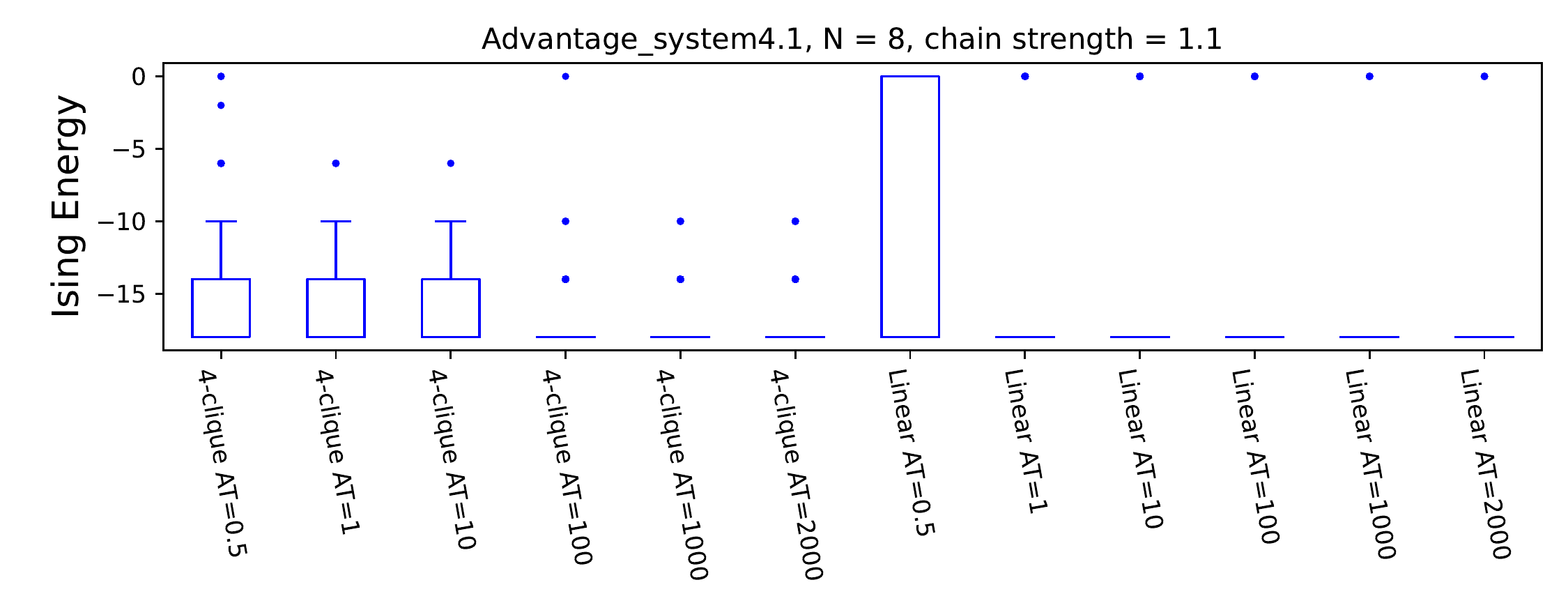}
    \includegraphics[width=0.49\textwidth]{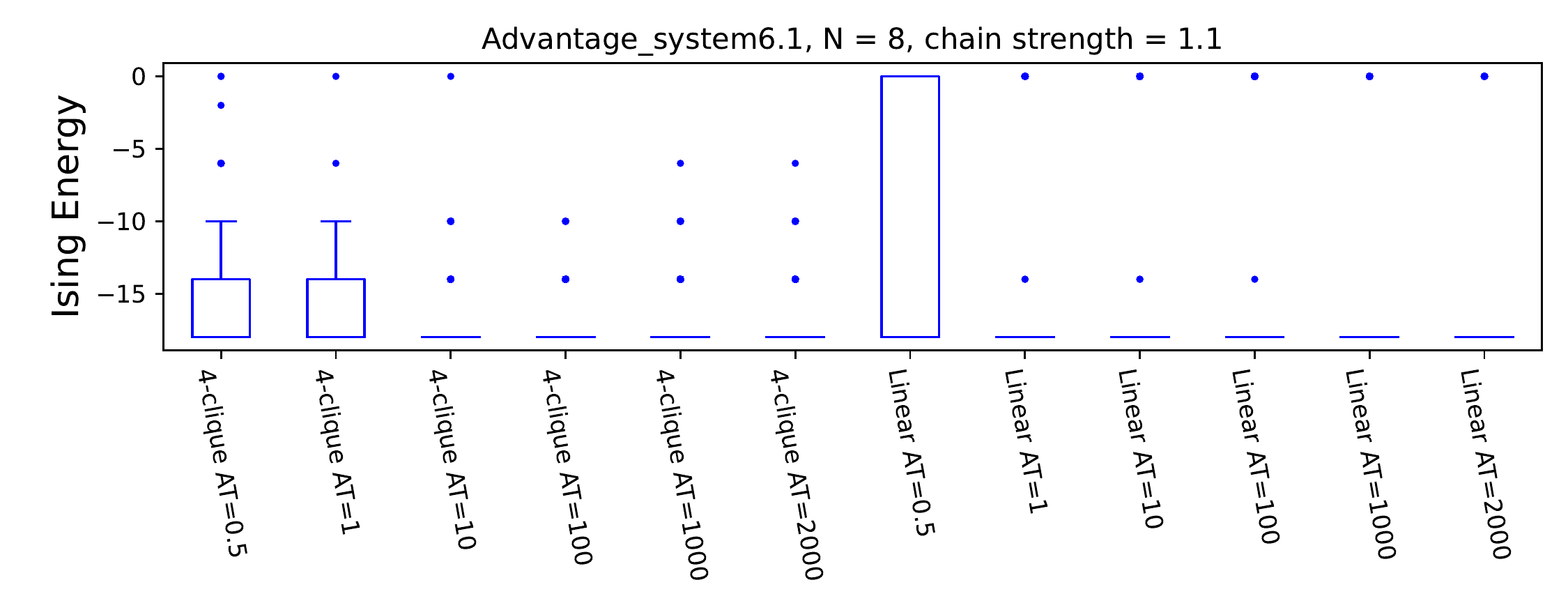}
    \caption{Sampling a single $N=8$ variable random spin glass instance using \texttt{Advantage\_system4.1} (left column), and \texttt{Advantage\_system6.1} (right column). Chain strength of $1.01$ (top row) and chain strength of $1.1$ (bottom row). Each plot is comprised of the spectrum of energy results from the 4-clique minor embedding in the left portion, and the corresponding equivalent linear path minor embedding energy spectrum plots in the right hand portion. Each set of data, for each combination of annealing time chain strength and minor embedding, is comprised of exactly $1000$ samples. Annealing times (AT) of $0.5$, $1$, $10$, $100$, $1000$, and $2000$ microseconds are used so as to evaluate how the two minor embeddings compare over different annealing times. Any outlier energy data points are represented as small blue dots, which may overlap on each other. }
    \label{fig:N8_results}
\end{figure*}

\subsection{Implementation on Quantum Annealing Hardware}
\label{section:methods_implementation}
An important point when implementing 4-clique minor embeddings is that the number of qubits used is likely significantly more than an equivalent linear embedding. Therefore, like when minor embedding very large problems with linear path chains, it is important to consider whether the uniform problem coefficient spreading causes the programmed weights to fall below the machine precision. If this is the case, for very large minor embeddings, it may be necessary to create non-uniform problem weight encoding on the chains. The number of couplers used to actually encode quadratic terms could also be varied; in the 4-clique embedding 4 couplers can be used, and those weights could be distributed in non-uniform ways (for example the weights could be placed entirely on one coupler). For the experimental results shown in Section~\ref{section:results}, uniform weight distributions are used for both the linear and the quadratic terms. 

Section~\ref{section:results} reports experimental energy results from using both the equivalent linear minor embedding and the 4-clique network minor embedding. The linear minor embedding is constructed from the 4-clique minor embedding by taking only one linear path down the 4-clique chain - and thereby using exactly one half of the qubits as the equivalent 4-clique minor embedding. This conversion from a 4-clique minor embedding to the linear path minor embedding is shown as side by side comparisons in Figures~\ref{fig:minor_embedding_N4} and~\ref{fig:minor_embedding_N8}. The chain strength is the primary parameter of interest when comparing these minor embeddings because that chain strength will be applied to both the linear path minor embeddings and the 4-clique minor embeddings. The relevant question is whether there is a difference between the two embeddings when the chain strength is the same. Importantly, the comparison against the linear path minor embeddings are a fair comparison of the impact of varying chain strength since the same region of the hardware graph is used. Note however that optimized linear path minor embeddings could have significantly shorter chain lengths and thus perform better for smaller problem instances where chain breaks and incoherent quantum annealing are dominant sources of error.

When executing the problem instances on the D-Wave quantum annealers, auto scaling is left on. Auto scaling is a backend parameter which, if left on, will scale the provided problem coefficients into the maximum energy scale that is possible on the quantum annealer, for both the linear and quadratic terms. Each parameter combination of the quantum annealer uses exactly $1000$ anneals. The annealing time is varied for the purpose of observing what effect it has on the results of the computation. The parameter \texttt{programming\_thermalization} was set to $0$ microseconds, and all other programming parameters are set to their default values. 

Another critical component of using minor embeddings is how to handle cases where the chains do not agree on the logical variable state (i.e. the chain is \emph{broken}). There are simple chain break resolution methods such as majority vote which can classically repair the chain with post processing. With the aim of illustrating the chain break frequency, along with solution quality in the D-Wave results reported in Section~\ref{section:results} no classical post processing is used. In particular, any sample which contains a broken chain is set to have an energy value of $0$, meaning that the overall measurement statistics still have $1000$ data points per setting. 

The problem instances we will consider are random spin glasses, defined on an all-to-all connected graph $G = (V, E)$ of size $N$;

\begin{equation}
    C(z) = \sum_{v \in V} w_v z_v + \sum_{(i,j) \in E} w_{ij} z_i z_j
    \label{equation:problem_instance}
\end{equation}

\begin{figure*}[t!]
    \centering
    \includegraphics[width=0.49\textwidth]{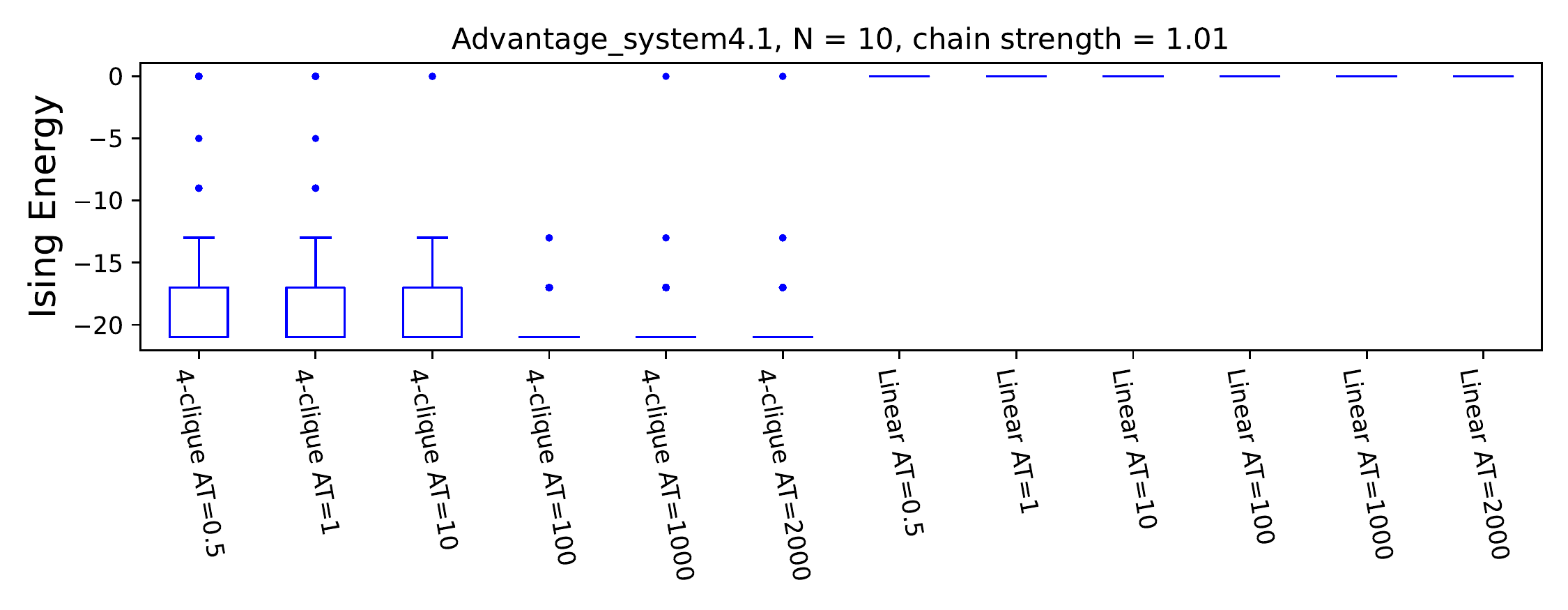}
    \includegraphics[width=0.49\textwidth]{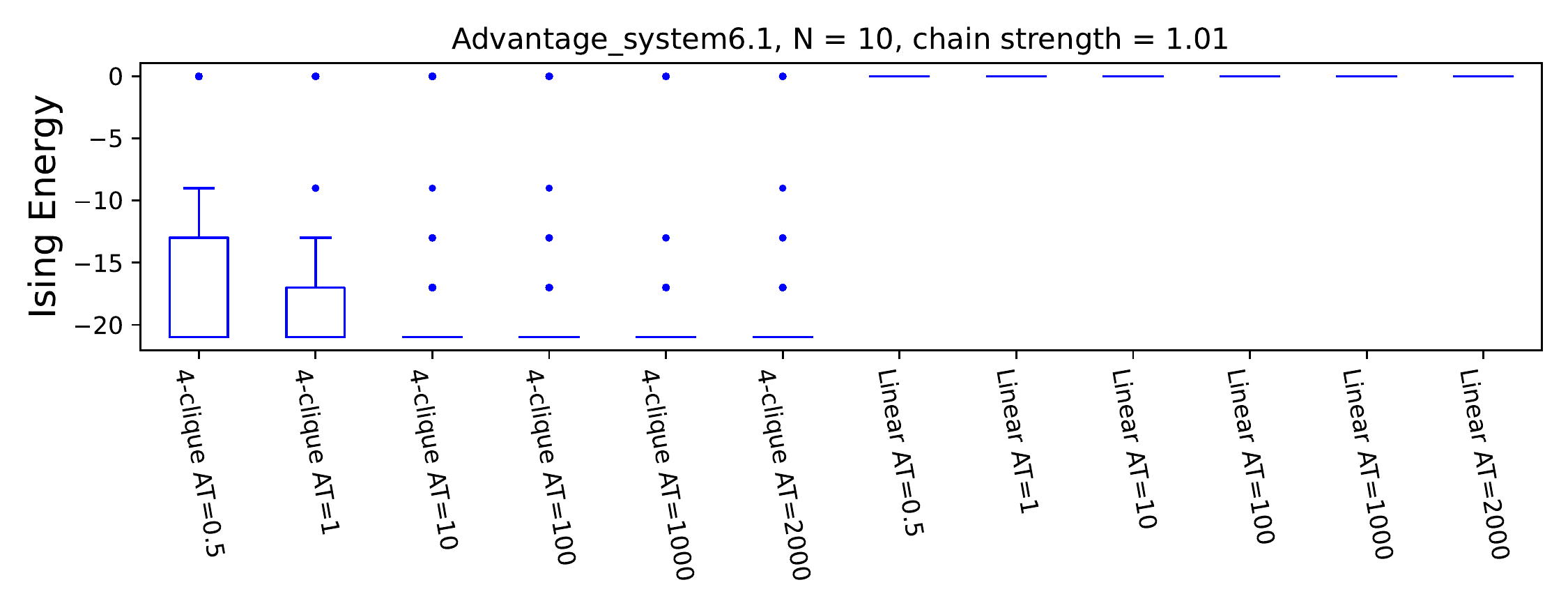}
    \includegraphics[width=0.49\textwidth]{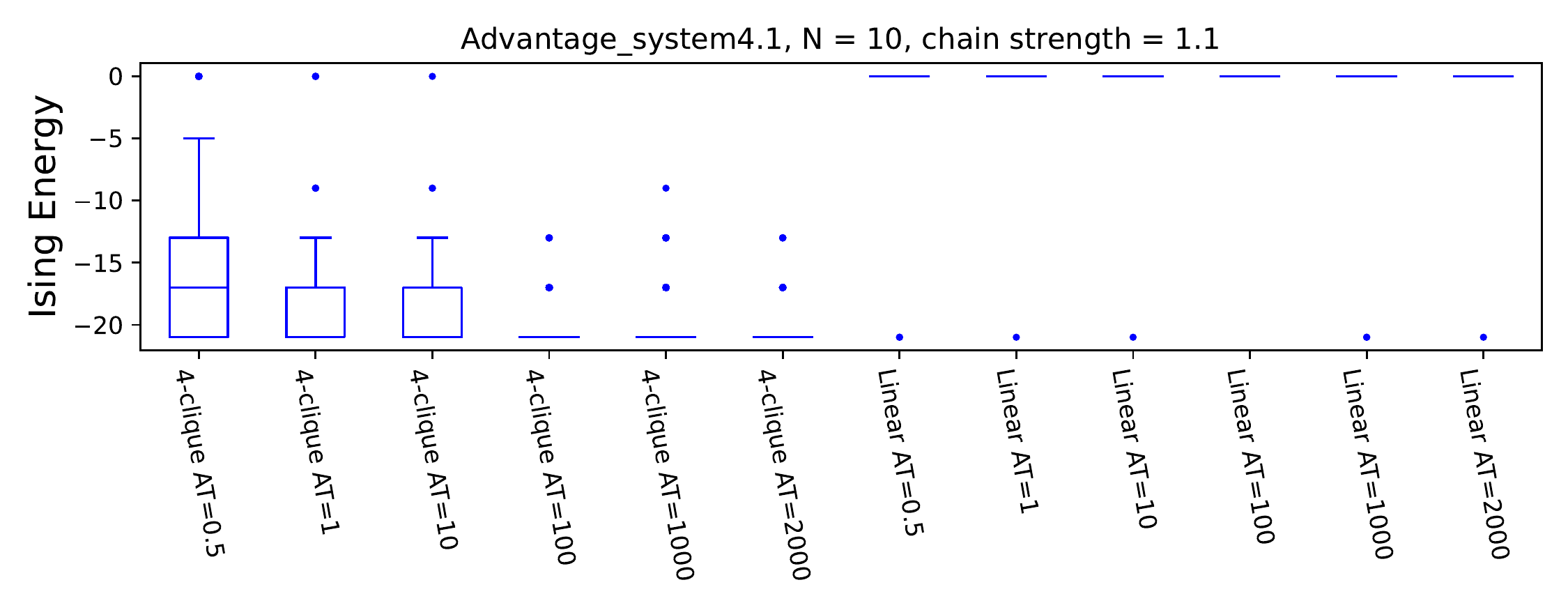}
    \includegraphics[width=0.49\textwidth]{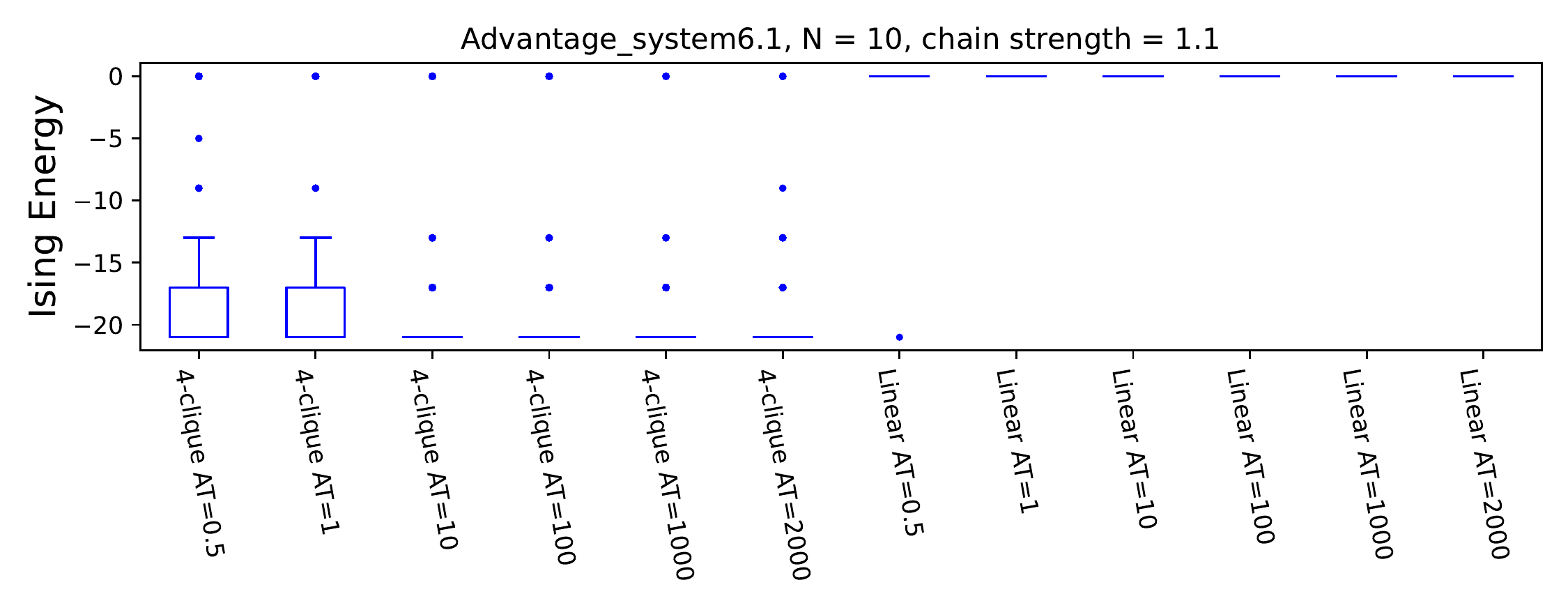}
    \caption{$N=10$ variable random spin glass sample energies from \texttt{Advantage\_system4.1} (left column), \texttt{Advantage\_system6.1} (right column). Chain strength of $1.01$ (top row) and chain strength of $1.1$ (bottom row). Each plot is comprised of the distribution of energy results from the 4-clique minor embedding in the left portion, and the corresponding equivalent linear path minor embedding energy spectrum plots in the right hand portion. Annealing times (AT) of $0.5$, $1$, $10$, $100$, $1000$, and $2000$ microseconds are used so as to evaluate how the two minor embeddings compare over different annealing times. }
    \label{fig:N10_results}
\end{figure*}

Where $w_{ij}$ and $w_v$ denote random coefficients, chosen uniformly at random from $\{+1, -1\}$, making these problems effectively discrete-coefficient Sherrington-Kirkpatrick models \cite{PhysRevLett.35.1792} with local fields. The goal is to find the vector of variables $z = [z_0, z_1, \dots, z_N]$ such that the cost function in eq.~\eqref{equation:problem_instance} is minimized, where the decision variables $z_i$ are spins ($+1$ or $-1$). This cost function evaluation, as it is an Ising model, is referred to as the \emph{energy} for that given set of variable assignments. For each problem size $N$ that is tested, a new problem instance is generated with new random coefficients.

\section{Results}
\label{section:results}

This section analyzes experimental results from executing random spin glasses on \texttt{Advantage\_system4.1} and \texttt{Advantage\_system6.1}.

Figure~\ref{fig:N8_results} shows a side by side comparison of 4-clique and linear path minor embeddings on $N=8$ random problem instances, using relatively small chain strengths of $1.1$ and $1.01$. Figure~\ref{fig:N8_results} shows that there is very little difference between the energy distributions for the two minor embeddings - both minor embeddings have very low chain break rates and both have converged to the same optimal solution. Note that this proof of concept with this $8$ variable problem instance would perform better on the hardware if encoded using optimized small chain length linear path minor embeddings compared to the 4-clique network minor embedding given how small the problem instance is. We will probe larger problem instances next so as to see how more complicated 4-clique minor embeddings perform.

\begin{figure*}[t!]
    \centering
    \includegraphics[width=0.49\textwidth]{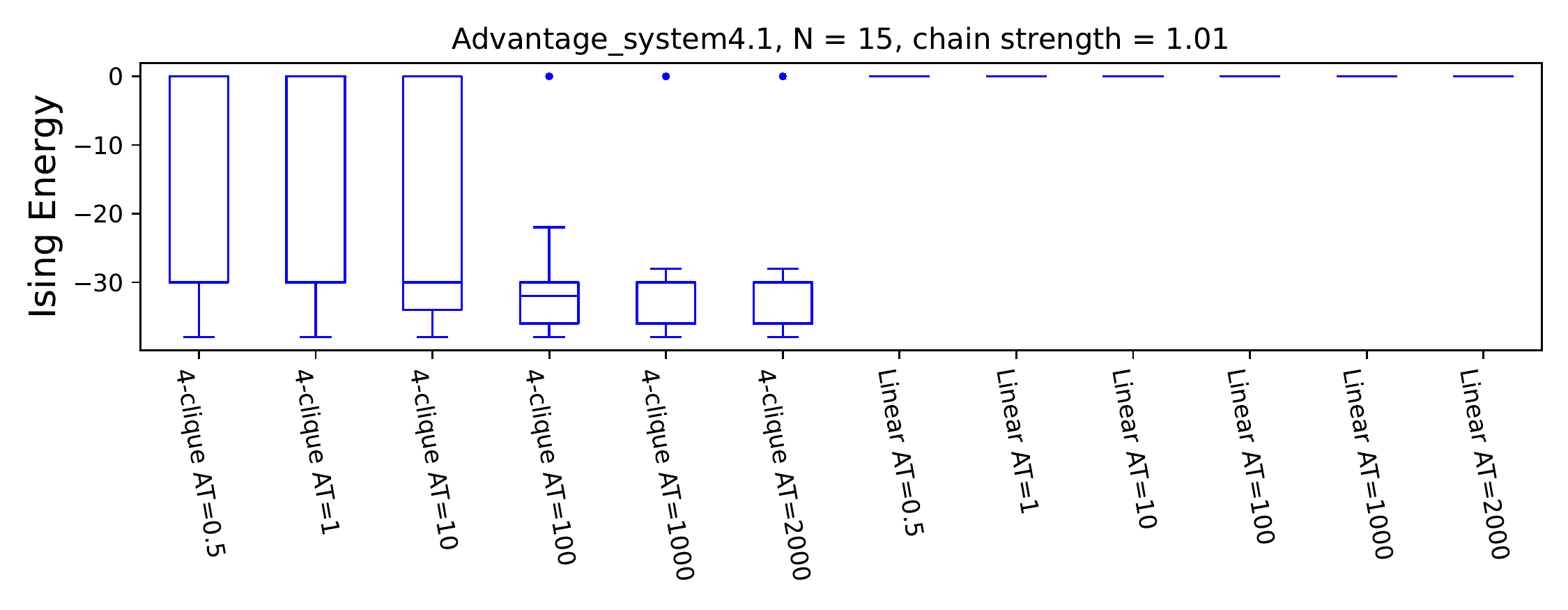}
    \includegraphics[width=0.49\textwidth]{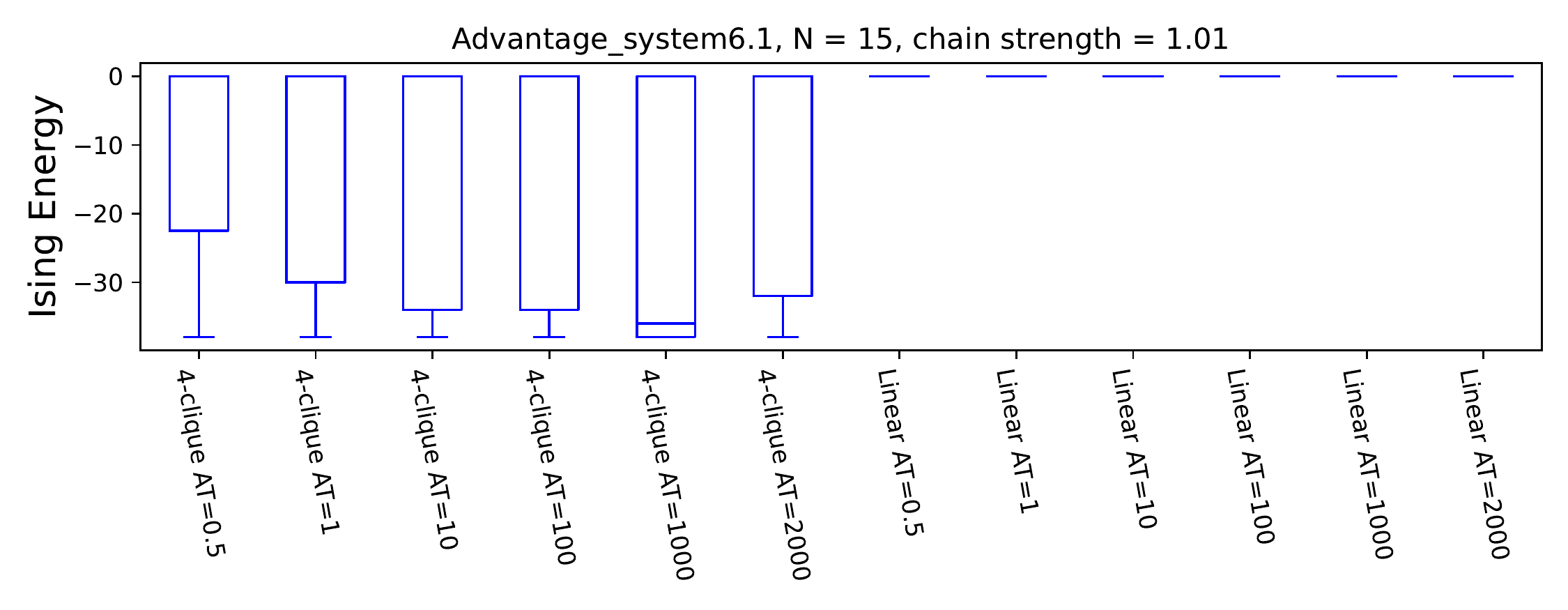}
    \includegraphics[width=0.49\textwidth]{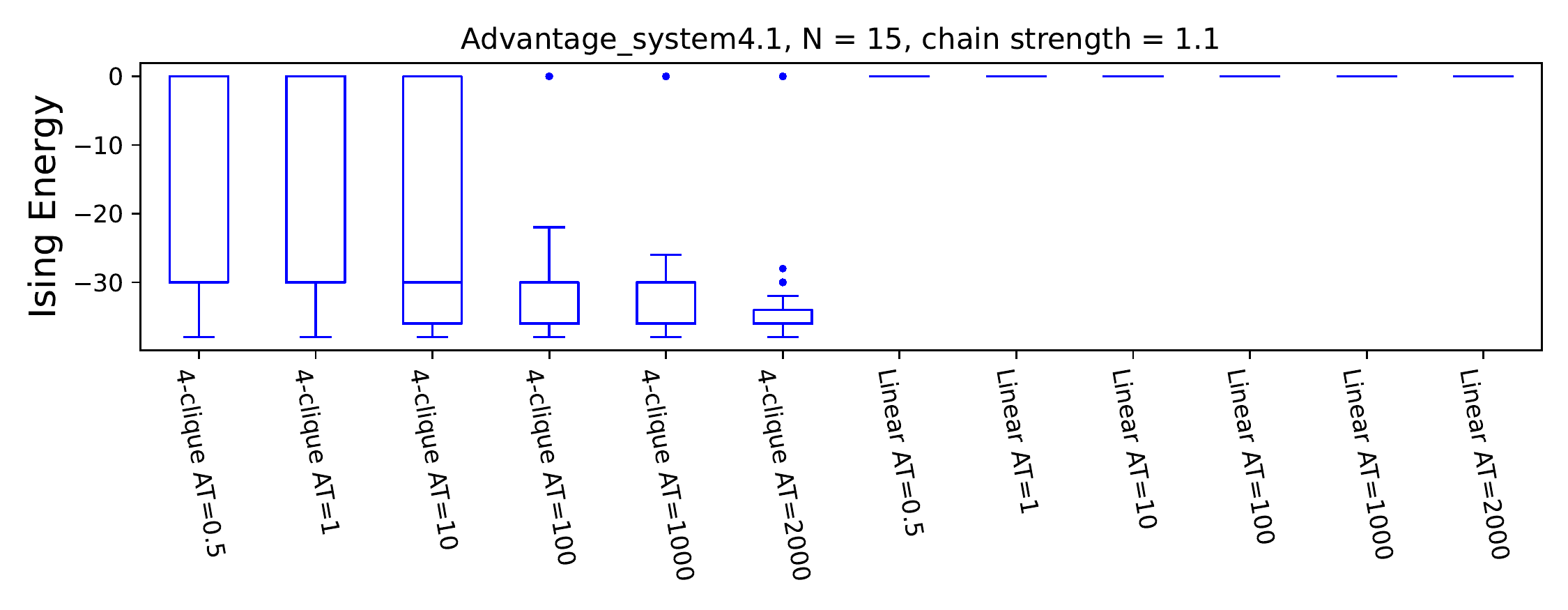}
    \includegraphics[width=0.49\textwidth]{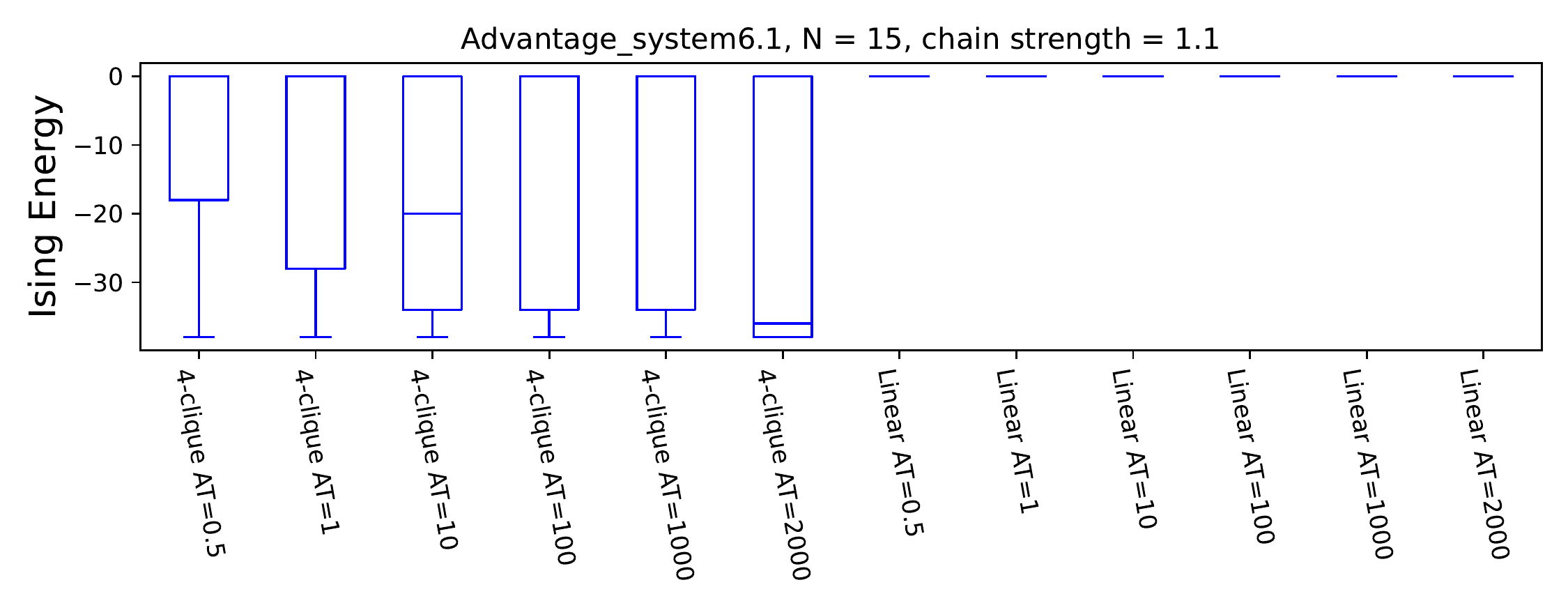}
    \caption{$N=15$ variable random spin glass energy distributions from \texttt{Advantage\_system4.1} (left column), \texttt{Advantage\_system6.1} (right column). Chain strength of $1.01$ (top row) and chain strength of $1.1$ (bottom row). Each plot is comprised of the spectrum of energy results from the 4-clique minor embedding in the left portion, and the corresponding equivalent linear path minor embedding energy spectrum plots in the right hand portion. Annealing times (AT) of $0.5$, $1$, $10$, $100$, $1000$, and $2000$ microseconds are used so as to evaluate how the two minor embeddings compare for different annealing times.  }
    \label{fig:N15_results}
\end{figure*}

Figure~\ref{fig:N10_results} shows the same comparison as Figure~\ref{fig:N8_results}, except the problem size was increased to $N=10$. In these plots, there is now a clear difference between the 4-clique and the comparable linear path minor embeddings. At these very low chain strengths, the greater connectivity of the 4-clique path minor embedding allowed the computation to remain stable and finds low energy solutions. By contrast, the linear path minor embedding has an extremely high chain break frequency and therefore the computations are not as robust at finding low energy solutions. This shows the 4-clique minor embedding is able to utilize a smaller proportion of the available programmable coefficient range on the hardware by being able to carry out the computation with only a chain strength of $1.01$ (note that the hardware option of autoscaling is turned on), compared the the equivalent minor embedding technique. There are a few outlier instances where the linear path minor embedding is able to sample the optimal solution, but only when the chain strength was set to $1.1$.

Figure~\ref{fig:N15_results} continues the trend observed in Figure~\ref{fig:N10_results} where the 4-clique minor embedding is able to sample low energy solutions with minimal chain breaks compared to the equivalent linear path minor embedding for a $N=15$ problem instance. Notably, all of the samples for the linear path minor embedding had broken chains and therefore clearly performed worse than the 4-clique minor embedding. This is a critical observation that the 4-clique network minor embedding was able to obtain low energy samples of the Ising model using the comparatively extremely small chain strength of $1.01$, leaving much of the available programmable coupler energy scale available for encoding of the coefficients of the original Ising model instead of dedicating that energy scale towards ferromagnetic chains. 

\begin{figure*}[t!]
    \centering
    \includegraphics[width=0.49\textwidth]{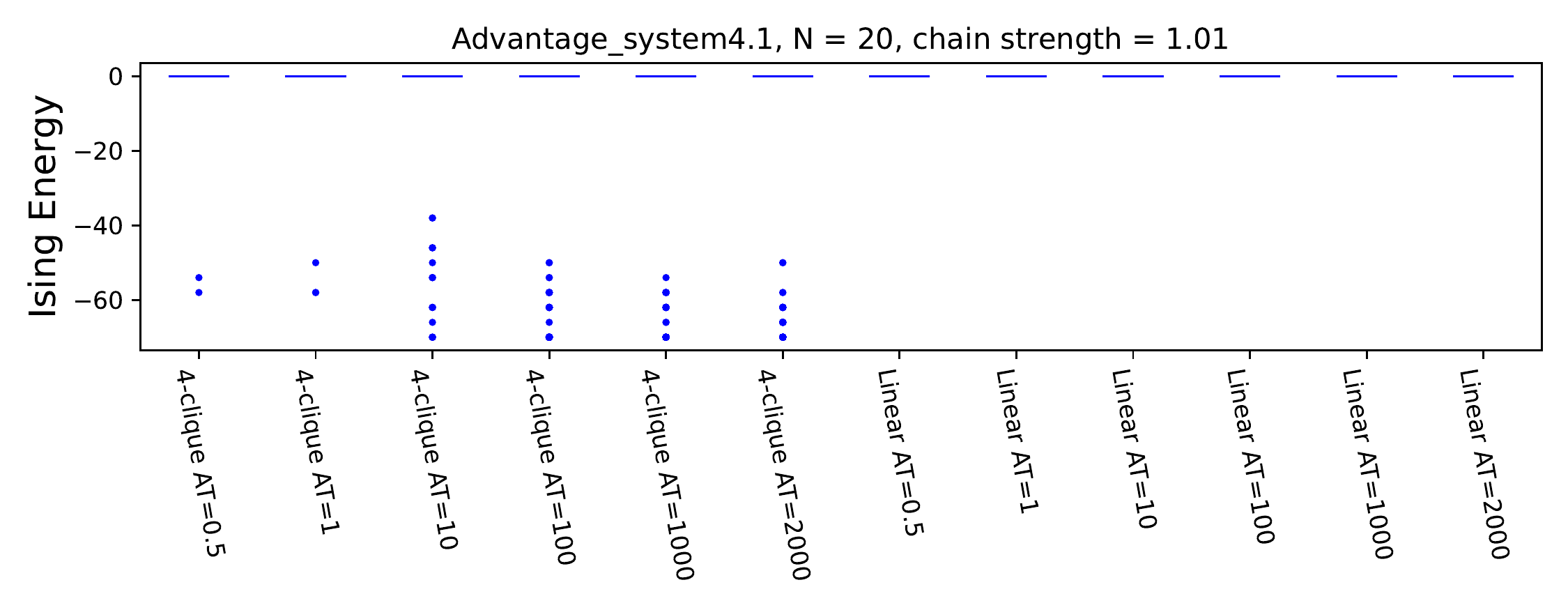}
    \includegraphics[width=0.49\textwidth]{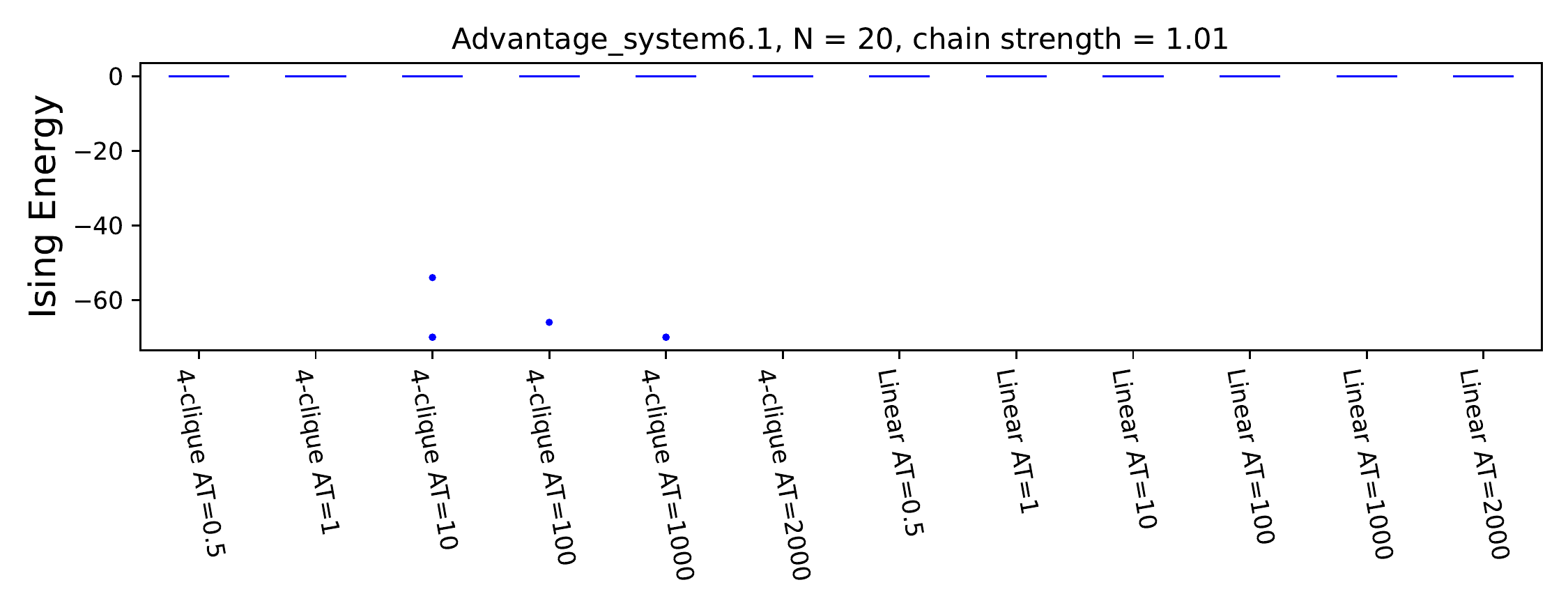}
    \includegraphics[width=0.49\textwidth]{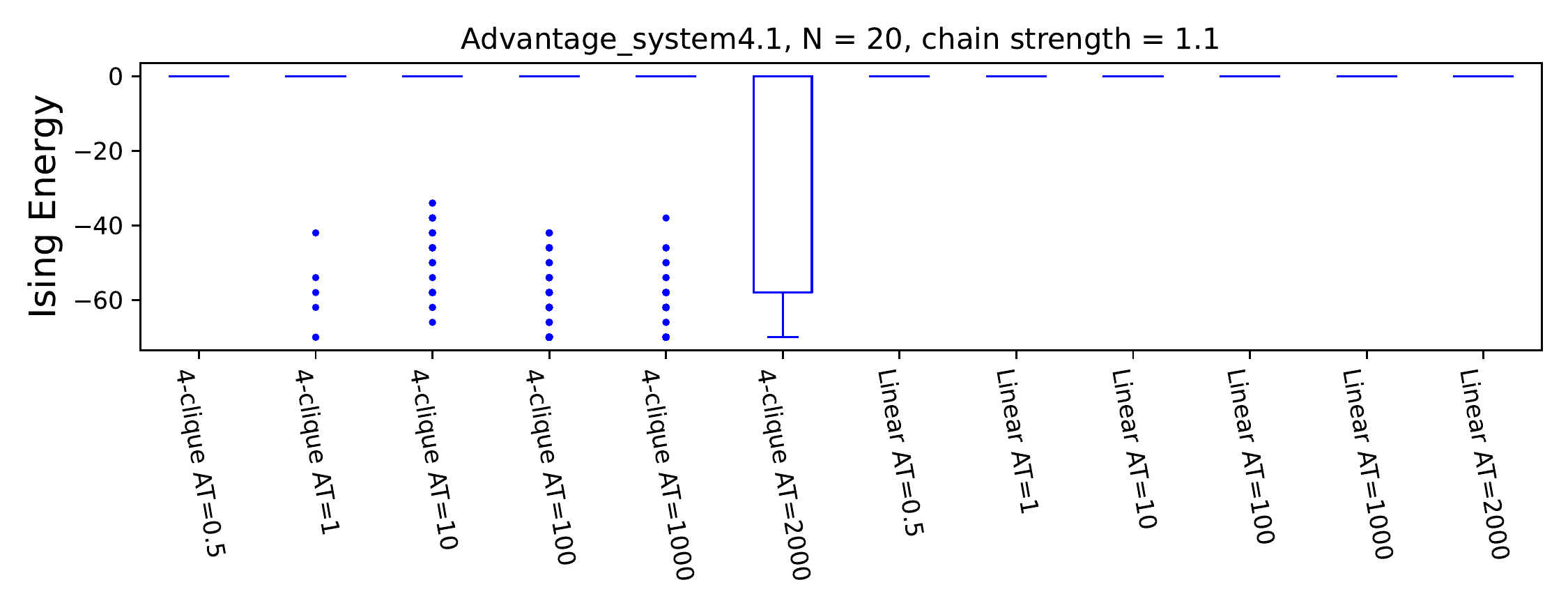}
    \includegraphics[width=0.49\textwidth]{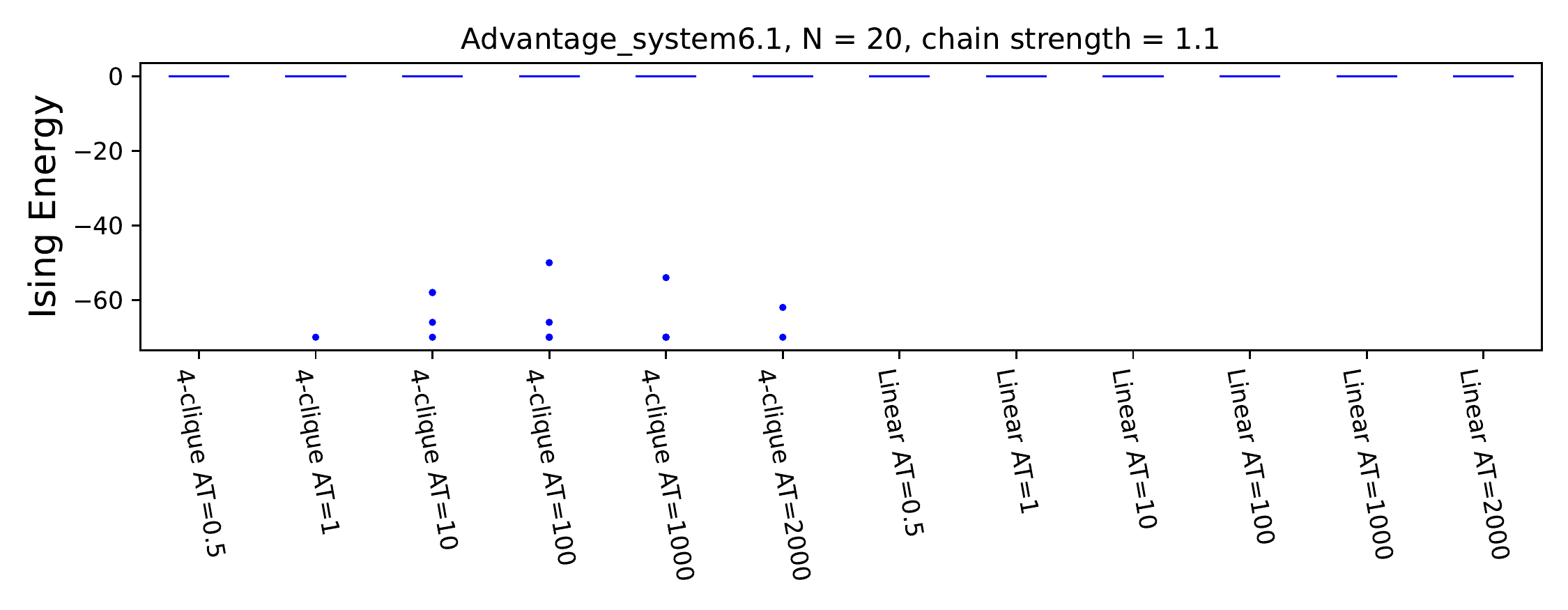}
    \includegraphics[width=0.49\textwidth]{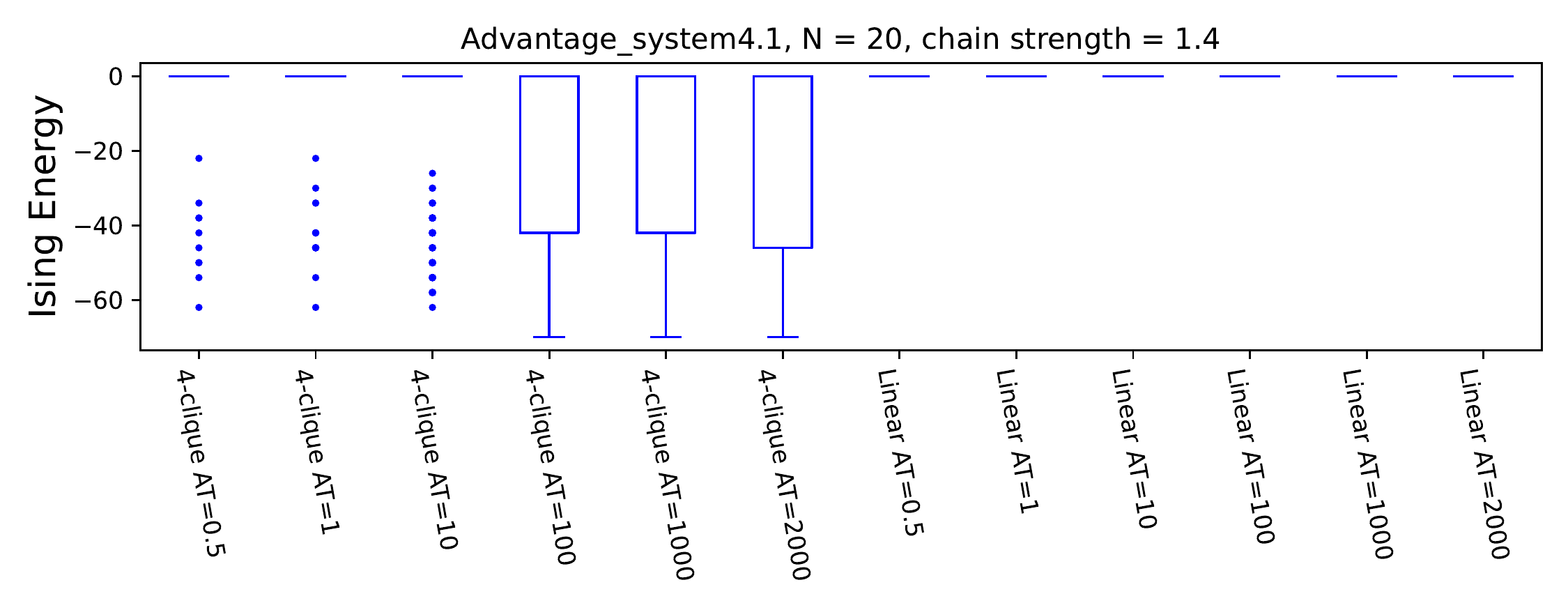}
    \includegraphics[width=0.49\textwidth]{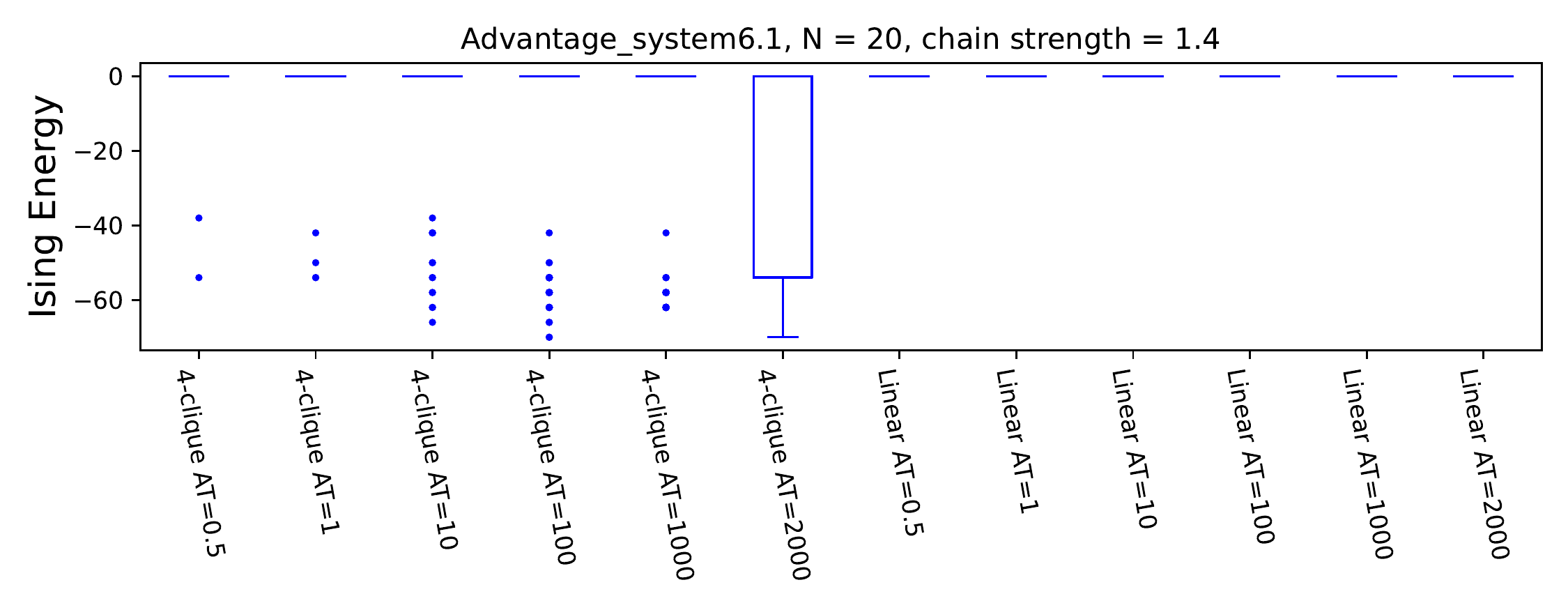}
    \caption{$N=20$ variable random spin glass sample energies from \texttt{Advantage\_system4.1} (left column), \texttt{Advantage\_system6.1} (right column). Chain strength of $1.01$ (top row), chain strength of $1.1$ (middle row), and chain strength of $1.4$ (bottom row). Each plot is comprised of the distribution of energy results from the 4-clique minor embedding in the left portion of the plot, and the corresponding equivalent linear path minor embedding energy spectrum plots in the right hand portion of the plot. Annealing times (AT) of $0.5$, $1$, $10$, $100$, $1000$, and $2000$ microseconds are used so as to evaluate how the two minor embeddings compare over different annealing times.  }
    \label{fig:N20_results}
\end{figure*}

Figure~\ref{fig:N20_results} shows the energy results for a $N=20$ problem instance. Because of the increased chain lengths (see Table~\ref{table:minor_embedding_chain_lengths}), the small chain strengths of $1.01$ and $1.1$ do not work well for this problem size. Therefore, this plot also includes results for a chain strength of $1.4$. Although the chain strength needed to be increased to see good low energy state sampling with the 4-clique minor embedding, it is still the case that the 4-clique embedding performed better than the linear path minor embedding especially with respect to chain break frequency. At this very long chain size, the linear path minor embedding results always had samples with broken chains.

\begin{figure*}[t!]
    \centering
    \includegraphics[width=0.49\textwidth]{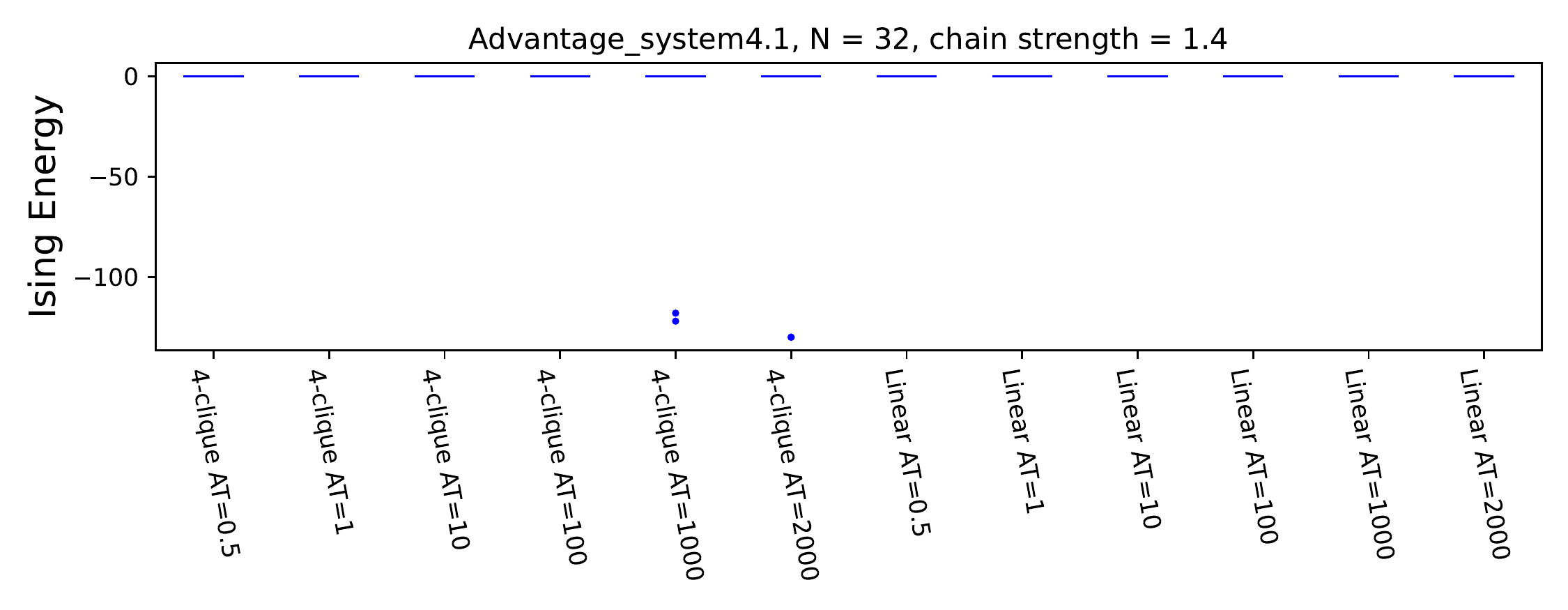}
    \includegraphics[width=0.49\textwidth]{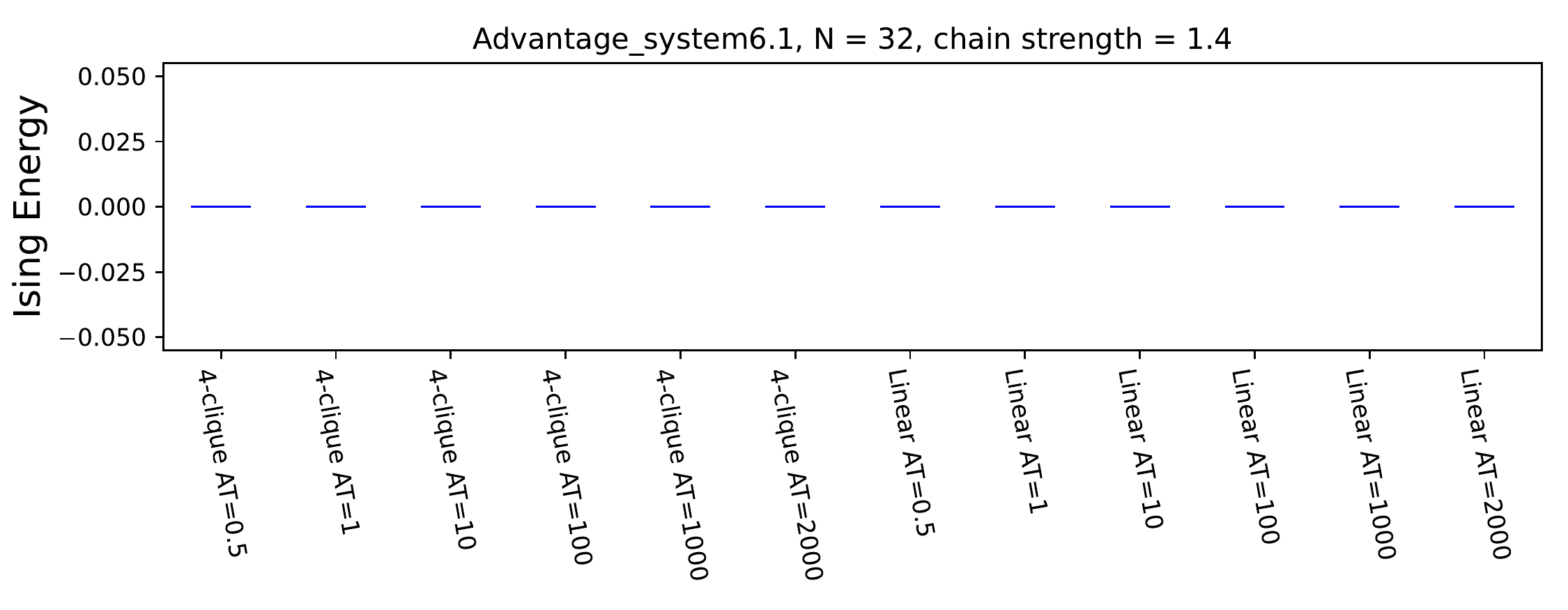}
    \includegraphics[width=0.49\textwidth]{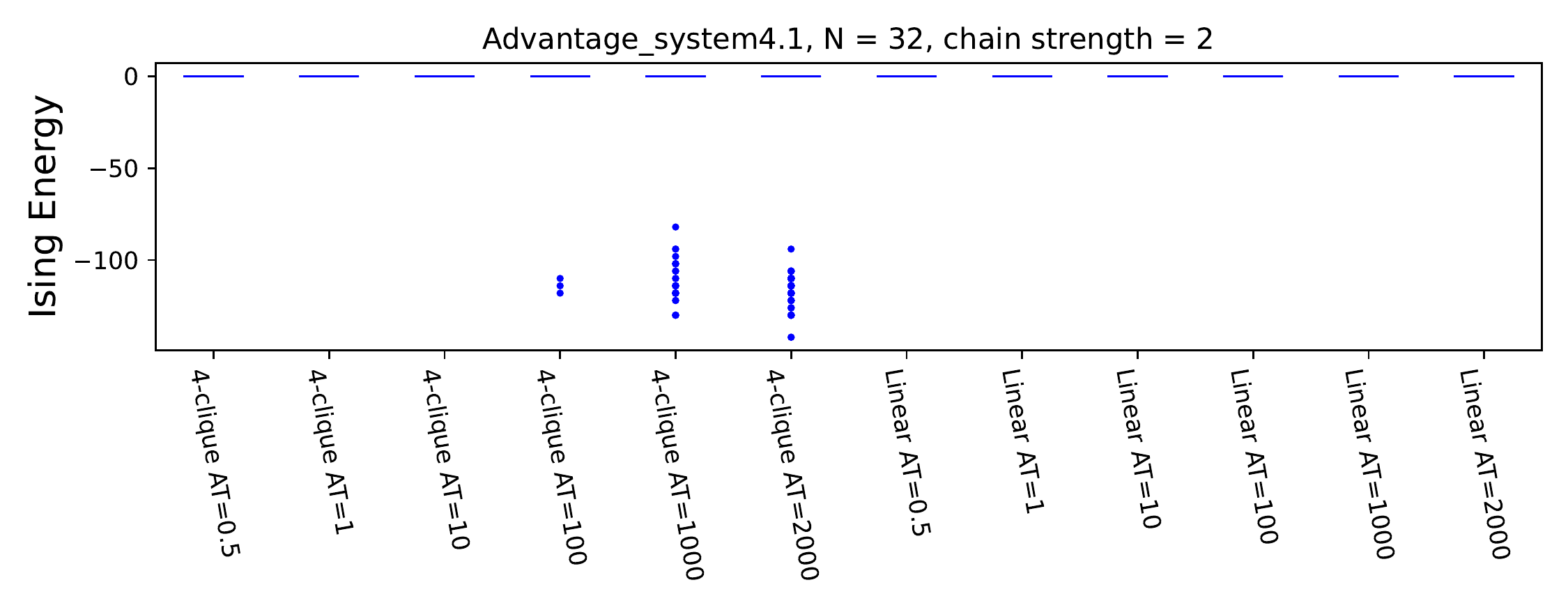}
    \includegraphics[width=0.49\textwidth]{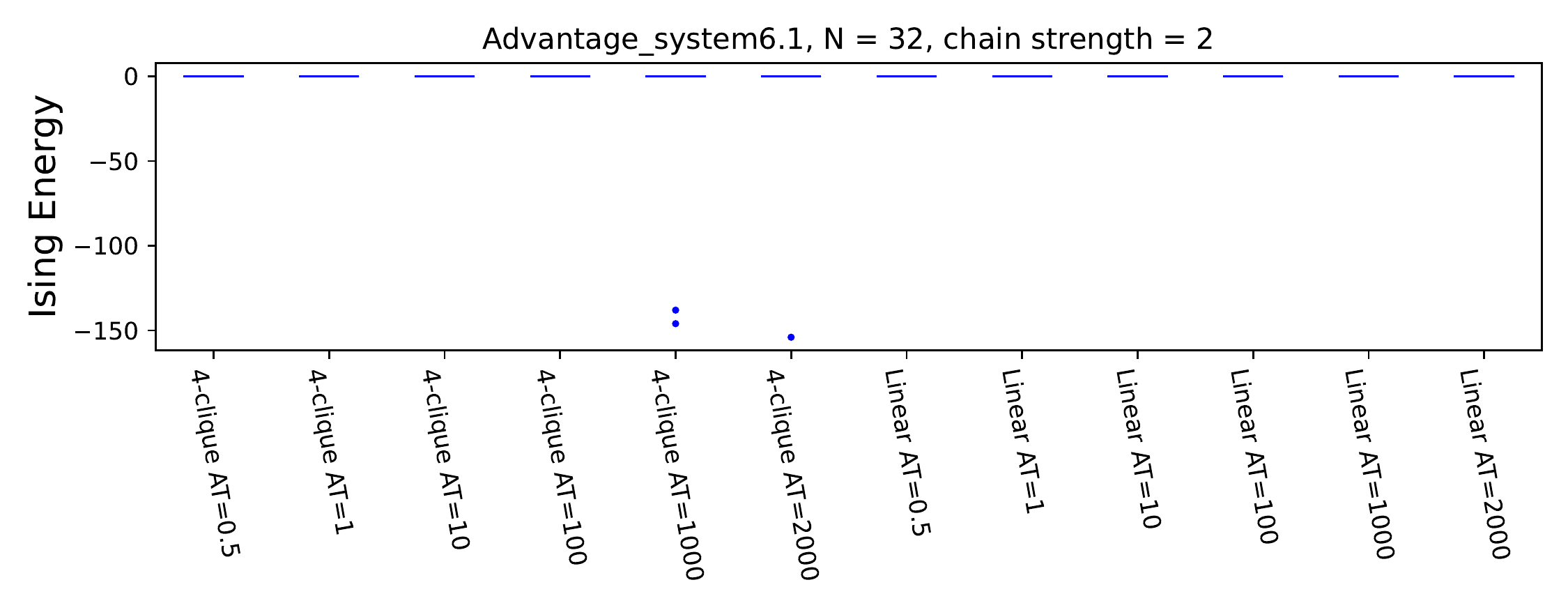}
    \includegraphics[width=0.49\textwidth]{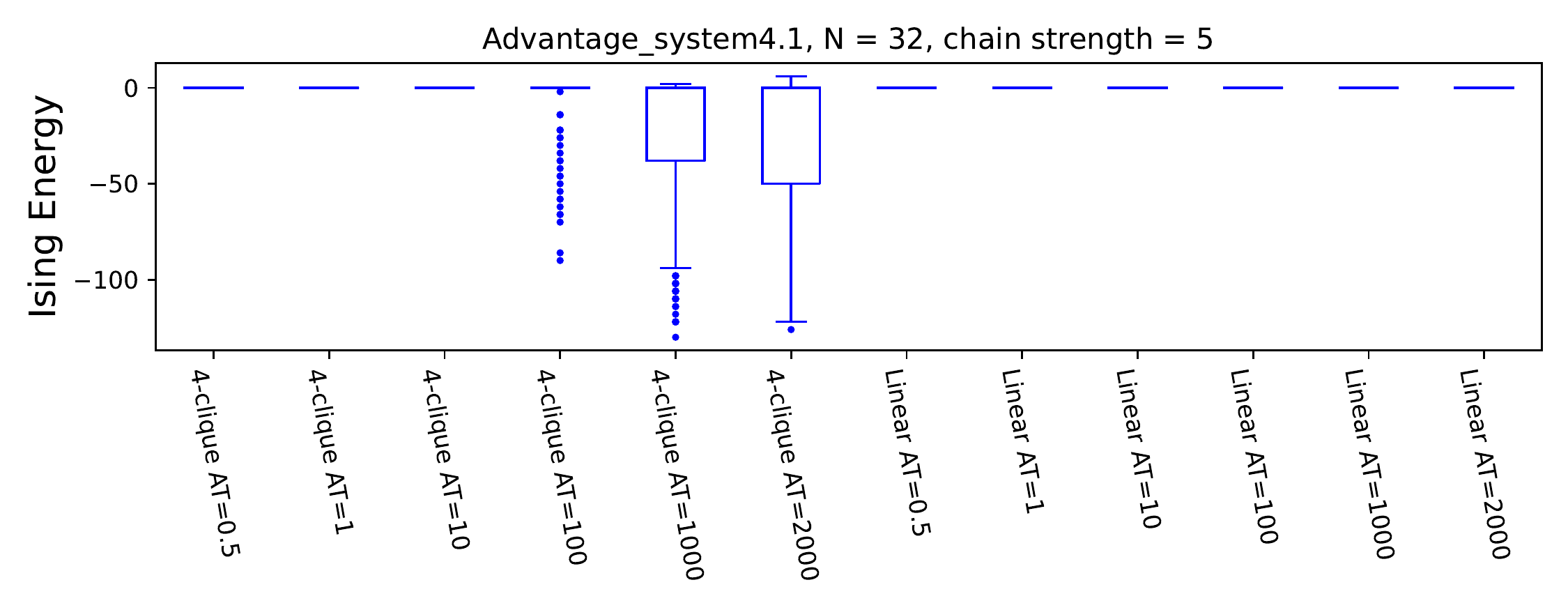}
    \includegraphics[width=0.49\textwidth]{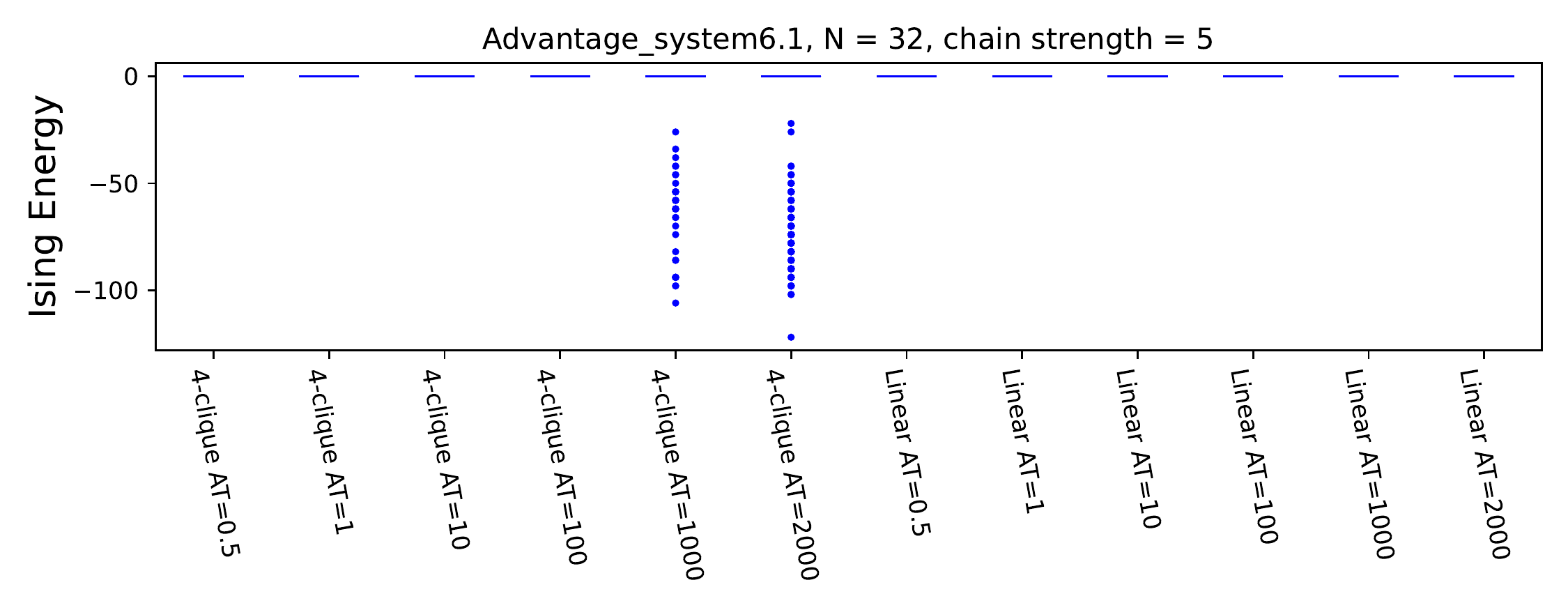}
    \includegraphics[width=0.49\textwidth]{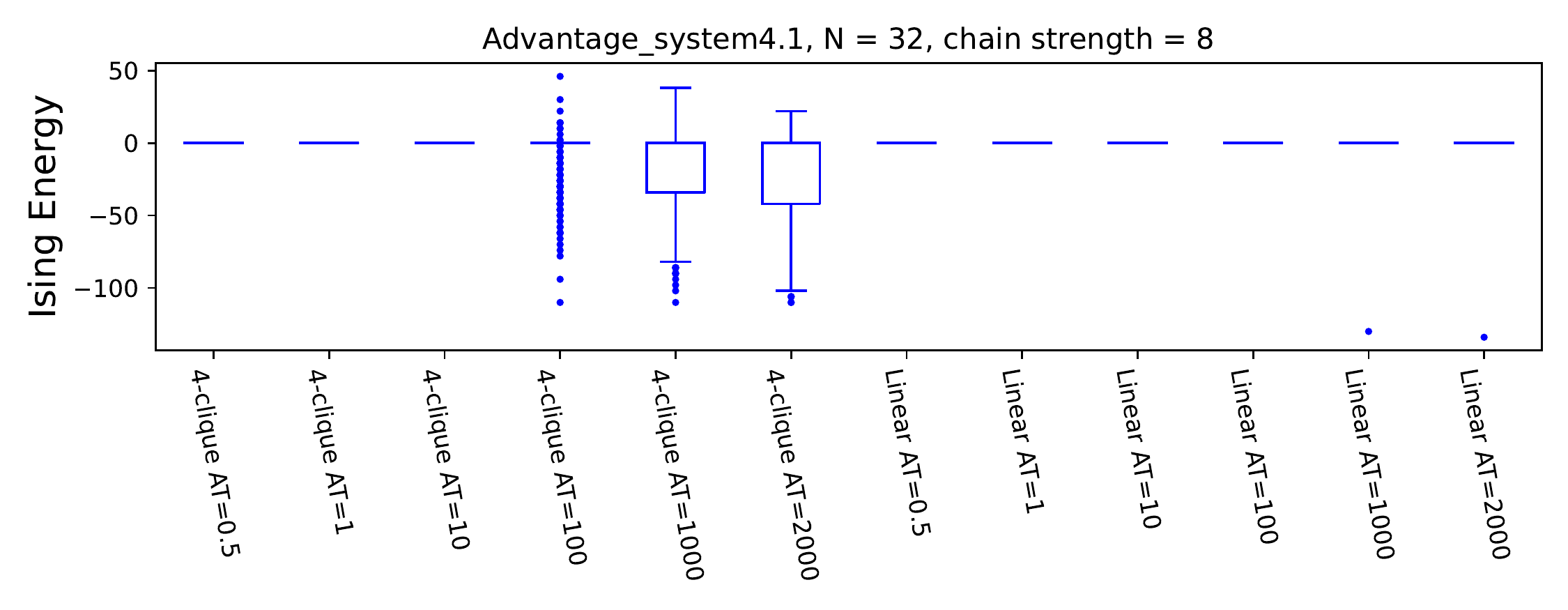}
    \includegraphics[width=0.49\textwidth]{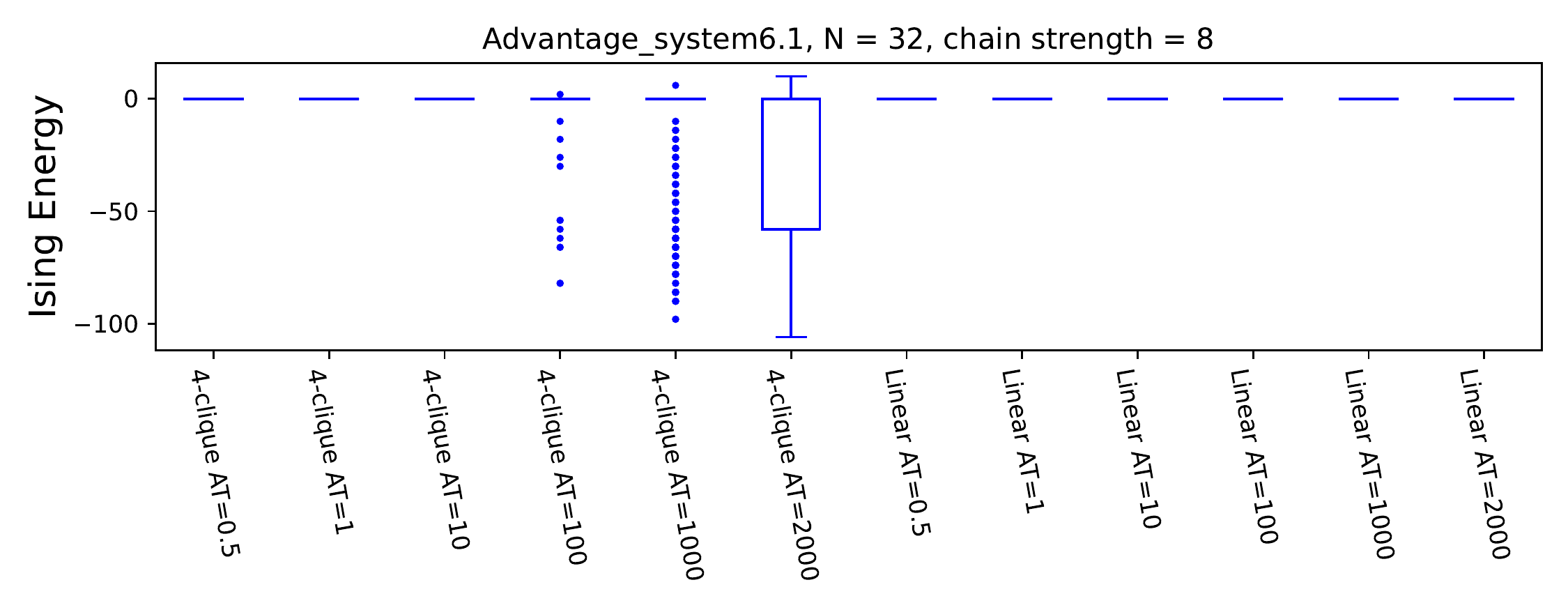}
    \caption{$N=32$ variable random spin glass energy results from \texttt{Advantage\_system4.1} (left column), \texttt{Advantage\_system6.1} (right column). Chain strength of $1.4$ (top row), chain strength of $2$ (top-middle row), chain strength of $5$ (middle-bottom row), chain strength of $8$ (bottom row). Each plot is comprised of the spectrum of energy results, represented by separate boxplots, from the 4-clique minor embedding in the left portion, and the corresponding equivalent linear path minor embedding energy spectrum plots in the right hand portion. Annealing times (AT) of $0.5$, $1$, $10$, $100$, $1000$, and $2000$ microseconds are used so as to evaluate how the two minor embeddings compare over different annealing times. }
    \label{fig:N32_results}
\end{figure*}

Figure~\ref{fig:N32_results} shows the energy results for a $N=32$ Ising model problem instance. The chain lengths for this problem size are dramatically larger than the all of the other tested minor embeddings (see Table~\ref{table:minor_embedding_chain_lengths}), and consequently much larger chain strengths were required to obtain reasonable results. At a chain strength of $1.4$, nearly all of the samples had broken chains or an energy of $0$; only the 4-clique minor embedding at $1000$ and $2000$ microsecond anneal times produced a few low energy samples. Once again, even at these larger chain strengths, the 4-clique minor embedding still had much fewer chain breaks compared to the linear path minor embedding and therefore better samples, at least at sufficiently long annealing times (e.g. $1000$ or $2000$ microseconds). However, here we begin to see where too large of chain strengths can be detrimental for the 4-clique minor embedding. At a chain strength of $8$, the 4-clique energy spectrum begins to clearly get worse than $0$, whereas with a chain strength of $5$ the energy results were actually better than with a chain strength of $8$. This shows that the chain strength used in the 4-clique minor embedding needs to be carefully tuned such that results do not get worse because of the chain strength using too much of the available programmable coupler weight. At comparatively small chain strengths the 4-clique minor embedding performs very well.

Figures~\ref{fig:N8_results},~\ref{fig:N10_results},~\ref{fig:N15_results},~\ref{fig:N20_results}, and~\ref{fig:N32_results} all show there is a clear trend for the 4-clique minor embedding across annealing times; the longer annealing times result in consistently lower energy solutions. These figures also all show that \texttt{Advantage\_system4.1} samples lower energy solutions more consistently compared to \texttt{Advantage\_system6.1}.

\section{Discussion and Conclusion}
\label{section:conclusion}

In this article we have introduced a new method for minor embedding discrete combinatorial optimization problems onto D-Wave quantum annealers with hardware graphs that contain networks of 4-cliques. Importantly, we do not expect this method to likely be used for the current $P_{16}$ Pegasus devices. The reason for this is that the largest minor embeddings which can be formed from the contracted 4-clique networks of $P_{16}$ graphs are not especially large in comparison to equivalent linear path minor embeddings which can be computed, and in particular linear path minor embeddings have considerably shorter chain lengths for small problem sizes and will thus perform considerably better and use less of the hardware graph than the proposed 4-clique network minor embedding. The reason for this is because of the longer chain lengths required to create a minor embedding on the contracted 4-clique graphs; linear path minor embeddings can be constructed to use much shorter chain lengths than the side by side comparisons done in Section~\ref{section:results}. For example, for problem instances with a small number of variables (such as on the order of $10$ variables) and a standard coefficient precision range, the standard optimized linear path minor embeddings will perform much better than a 4-clique network minor embedding on a Pegasus graph due to decreased hardware usage and likely more coherent quantum annealing. The regime where the 4-clique network minor embedding could be potentially useful is for much larger quantum annealers (which can form a 4-clique network, such as devices with a Pegasus hardware graph) where minor embedding of significantly large problem sizes (e.g. hundreds or thousands of logical variables) will require exceedingly large ferromagnetic chain strengths and long chains using the equivalent linear path chain minor embedding (which, if the chains are sufficiently long, will be prone to breaking). There, the 4-clique chains could provide lower levels of chain breaks due to the increased inter-chain integrity and require a smaller proportion of the programmable energy scale of the quantum annealer compared to alternative minor embedding methods. For example, if for a fully connected problem instance embedded on a large sparse hardware graph the linear path chain lengths are already significantly long such that significant chain breaks occur (and the maximum chain strengths are limited), then the 4-clique network minor embedding could be applied in order to perform improved quantum annealing on that hardware platform. On sparse hardware graphs, fully connected minor embeddings quickly use the available hardware graph as the number of variables increases and thus use long chains. Therefore 4-clique network minor embeddings could be used most effectively for fully connected minor embeddings on future large quantum annealing hardware graphs where long chains are inevitable even for linear path minor embeddings. Because of the limited size of current quantum annealing hardware, this type of large problem instance minor embedding comparison can not be performed on current D-Wave quantum annealers but this comparison should be performed on future quantum annealing hardware.

Additionally, even for small problem sizes, if the energy scale range of the problem coefficients is incredibly important for a minor embedded problem, then the 4-clique network minor embedding could be employed to make more of the coefficient range available on the chip to be used for problem coefficients, instead of inter-chain couplers. This type of problem may occur for optimization problems where the high precision of the coefficients being programmed onto the hardware graph is important to represent the problem faithfully compared to what can be programmed using linear path minor embeddings, which can occur if the coefficient range of the original problem is large. 

There are several future research questions that can be considered:

\begin{enumerate}[noitemsep]
    \item Make the weights on the chain non-uniform - perhaps make them proportional to their degree within the 4-clique chain.
    \item Use flux bias offset and anneal offset calibration~\cite{chern2023tutorial, DWave_err_corr} to balance the chain statistics in the minor embeddings, both for 4-clique and the equivalent linear minor embeddings, with the goal of reducing bias in the quantum annealing samples.
    \item Create more structured minor embeddings of the contracted 4-clique graphs, specifically with the goal of making more uniform chain lengths in the minor embeddings.
    \item Investigate encoding variable states across even larger subgraphs of the hardware topology. By encoding variable states into larger pieces of the physical hardware, it becomes harder for noise to induce errors in the physical group of qubits. 
    \item Investigate how the 4-clique chains break. Is there a pattern with respect to the qubit degree within the 4-clique chain, or with respect to the position of the qubit in the chain?
    \item Numerical simulations of exact quantum state evolution when using the standard linear minor embedding compared to the proposed 4-clique network minor embedding. Currently, such classical simulators are severely limited in the total number of qubits that can be simulated, making comparisons between even smaller toy examples computationally difficult. Future improved quantum annealing simulation software could be used to perform such numerical simulations, giving better insights on how these two minor embedded methods work.
    \item Use denser minor-embeddings, such as the 4-clique network minor embedding, selectively and not for every single logical variable. For example, nodes in the logical graph which are highly connected could be represented by a large dense chain (such as a 4-clique chain), while less dense variables are represented by smaller chains. 
\end{enumerate}

\section{Acknowledgments}
\label{sec:acknowledgments}
This work was supported by the U.S. Department of Energy through the Los Alamos National Laboratory. Los Alamos National Laboratory is operated by Triad National Security, LLC, for the National Nuclear Security Administration of U.S. Department of Energy (Contract No. 89233218CNA000001). The research presented in this article was supported by the Laboratory Directed Research and Development program of Los Alamos National Laboratory under project numbers 20220656ER and 20210114ER and by the NNSA's Advanced Simulation and Computing Beyond Moore's Law Program. This research used resources provided by the Los Alamos National Laboratory Institutional Computing Program, which is supported by the U.S. Department of Energy National Nuclear Security Administration under Contract No. 89233218CNA000001. Thanks to Carleton Coffrin for helpful discussions. This work has been assigned the LANL technical report number LA-UR-23-20504.

\appendix

\section{Contracted 4-clique graphs on Zephyr}
\label{section:4_clique_zephyr}
\begin{figure*}[h!]
    \centering
    \includegraphics[width=0.49\textwidth]{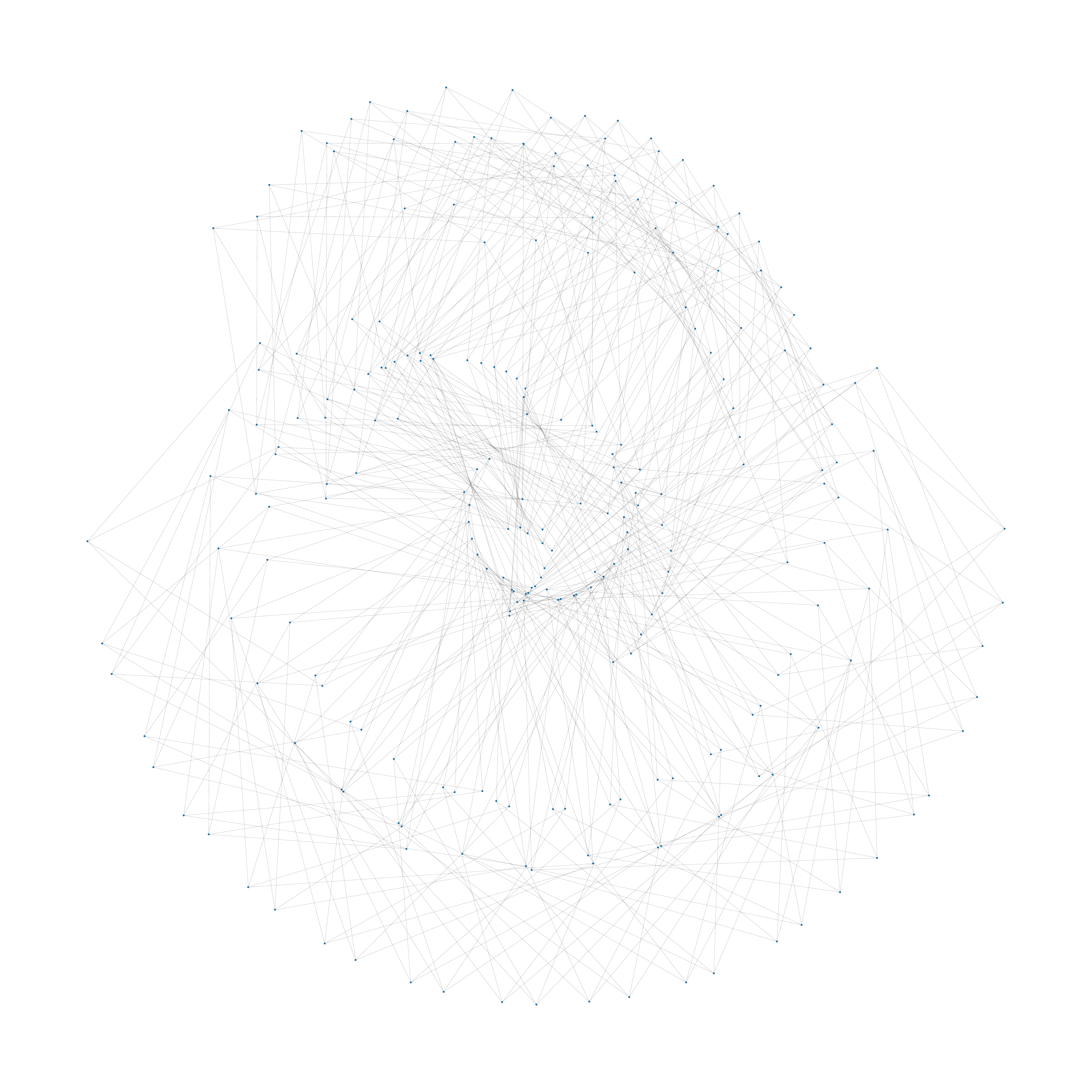}
    \includegraphics[width=0.49\textwidth]{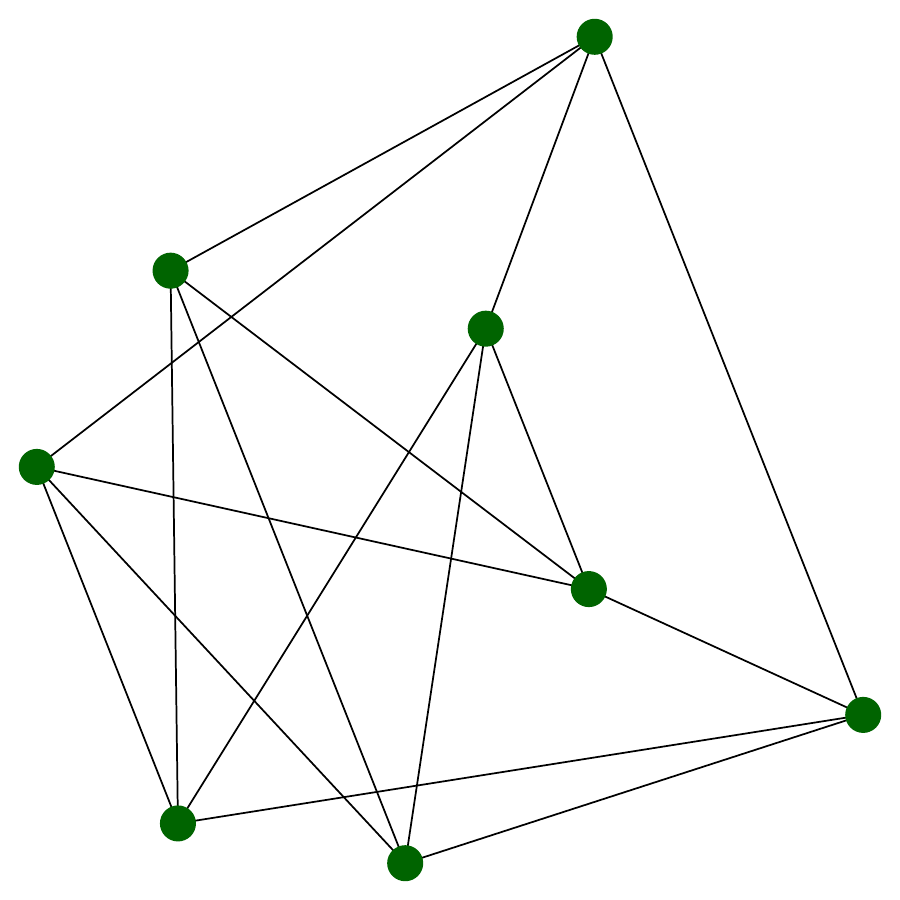}
    \caption{Contracted 4-clique graph for a Zephyr $Z_{16}$ graph (left), which is composed of $31$ unconnected graphs. Each of those $31$ unconnected subgraphs are isomorphic to each other; their structure is shown in the right side plot. These graphs are drawn using the kamada kawai layout algorithm. }
    \label{fig:zephyr_contracted_cliques}
\end{figure*}

Figure~\ref{fig:zephyr_contracted_cliques} shows the contracted 4-clique graphs for a logical Zephyr $Z_{16}$ graph with no hardware defects.

\section{Contracted 4-clique Pegasus minor embedding chain lengths}
\label{section:table_4_clique_chain_lengths}

Table~\ref{table:minor_embedding_chain_lengths} details chain length statistics of the computed contracted 4-clique random minor embeddings.

\begin{table*}[h!]
\begin{center}
\begin{tabular}{ |c||p{5.2cm}|p{5.2cm}| }
\hline
N & \texttt{Advantage\_system4.1} \newline 4-clique minor embedding chain lengths \newline (min, mean $\pm \sigma$, max) & \texttt{Advantage\_system6.1} \newline 4-clique minor embedding chain lengths \newline (min, mean $\pm \sigma$, max) \\ 
\hline
\hline
3 & (2, 2.667 $\pm$ 0.943, 4) & (2, 2.667 $\pm$ 0.943, 4) \\
\hline
4 & (2, 3.0 $\pm$ 1.0, 4)  & (2, 3.0 $\pm$ 1.0, 4) \\
\hline
5 & (2, 3.2 $\pm$ 0.98, 4) & (2, 3.2 $\pm$ 0.98, 4) \\
\hline
6 & (2, 4.0 $\pm$ 1.633, 6) & (2, 4.0 $\pm$ 1.633, 6) \\
\hline
7 & (2, 4.857 $\pm$ 2.1, 8) & (2, 4.857 $\pm$ 2.1, 8) \\
\hline
8 & (4, 7.0 $\pm$ 2.236, 10) & (4, 7.0 $\pm$ 2.646, 10) \\
\hline
9 & (4, 8.222 $\pm$ 2.393, 12) & (4, 8.222 $\pm$ 2.393, 12) \\
\hline
10 & (6, 10.4 $\pm$ 2.939, 14) & (6, 10.4 $\pm$ 2.939, 14) \\
\hline
11 & (8, 13.818 $\pm$ 3.459, 18) & (6, 12.182 $\pm$ 3.242, 18) \\
\hline
12 & (10, 15.667 $\pm$ 3.986, 22) & (8, 16.167 $\pm$ 6.189, 26) \\
\hline
13 & (12, 18.769 $\pm$ 4.475, 26) & (12, 19.231 $\pm$ 4.933, 28) \\
\hline
14 & (14, 23.429 $\pm$ 5.368, 30) & (12, 22.429 $\pm$ 5.716, 28) \\
\hline
15 & (22, 27.467 $\pm$ 3.222, 32) & (16, 25.867 $\pm$ 4.646, 34) \\
\hline
16 & (16, 31.75 $\pm$ 7.71, 42) & (18, 30.75 $\pm$ 7.137, 42) \\
\hline
17 & (24, 34.824 $\pm$ 5.576, 42) & (16, 35.647 $\pm$ 10.295, 46) \\
\hline
18 & (26, 37.778 $\pm$ 6.25, 46) & (24, 37.222 $\pm$ 7.634, 46) \\
\hline
19 & (26, 43.263 $\pm$ 7.9, 52) & (32, 41.579 $\pm$ 7.802, 56) \\
\hline
20 & (34, 47.7 $\pm$ 7.246, 58) & (28, 47.5 $\pm$ 11.897, 64) \\
\hline
21 & (36, 50.476 $\pm$ 7.998, 60) & (32, 50.095 $\pm$ 7.47, 60) \\
\hline
22 & (32, 54.455 $\pm$ 9.552, 66) & (32, 51.909 $\pm$ 11.305, 64) \\
\hline
23 & (40, 60.261 $\pm$ 13.835, 82) & (38, 56.087 $\pm$ 7.235, 64) \\
\hline
24 & (44, 62.417 $\pm$ 11.604, 84) & (34, 61.667 $\pm$ 13.972, 80) \\
\hline
25 & (44, 66.48 $\pm$ 9.753, 78) & (44, 63.92 $\pm$ 10.438, 82) \\
\hline
26 & (44, 70.462 $\pm$ 13.165, 92) & (52, 69.538 $\pm$ 9.548, 86) \\
\hline
27 & (54, 73.185 $\pm$ 9.495, 86) & (56, 73.778 $\pm$ 8.35 , 84) \\
\hline
28 & (60, 81.214 $\pm$ 13.356, 106) & (56, 79.5 $\pm$ 9.571 , 92) \\
\hline
29 & (60, 85.31 $\pm$ 12.879, 104) & (54, 77.655 $\pm$ 12.606 , 96) \\
\hline
30 & (66, 94.4 $\pm$ 16.584, 122) & (62, 92.867 $\pm$ 14.603 , 112) \\
\hline
31 & (70, 96.839 $\pm$ 18.815, 124) & (68, 94.645 $\pm$ 14.063, 116) \\
\hline
32 & (62, 99.625 $\pm$ 20.152, 134) & (70, 97.688 $\pm$ 20.689, 132) \\
\hline
\end{tabular}
\end{center}
\caption{Summary statistics for random 4-clique all-to-all minor embedding chain lengths that were computed using several iterations of the minorminer embedding heuristic. The all-to-all minor embeddings were computed for sizes $3$ through $32$, represented as each row in the table. Specifically, of the chain lengths in the minor embedding, the minimum, maximum, mean, and standard deviation of those lengths are reported for the minor embeddings computed on the contracted 4-clique graphs of \texttt{Advantage\_system4.1} and \texttt{Advantage\_system6.1}. Here chain length is referring to the total number of physical qubits used in the minor embedding. All quantities are rounded to three decimal places. }
\label{table:minor_embedding_chain_lengths}
\end{table*}

\clearpage
\setlength\bibitemsep{0pt}
\printbibliography

\end{document}